\documentclass[pmlr]{jmlr}

\RequirePackage{graphicx}
\usepackage{booktabs}
\usepackage{multirow}
\usepackage{enumitem}
\usepackage{microtype}
\usepackage{hyperref}
\usepackage{xurl}

\jmlrvolume{340} 
\jmlryear{2026}
\jmlrworkshop{Machine Learning for Healthcare}
\jmlrproceedings{PMLR}{Proceedings of Machine Learning Research}

\title[The Cost of Binarizing Survival Outcomes in Clinical Prognostic Modeling]{The Cost of Binarizing Survival Outcomes in Clinical Prognostic Modeling}

\author{\Name{Shashank Yadav}
       \Email{yadav.shashank@mayo.edu}\\ 
       \addr Department of Radiation Oncology\\
       Mayo Clinic, Rochester, MN, USA 55905 
       \AND
       \Name{David M. Routman}
       \Email{routman.david@mayo.edu}\\ 
       \addr Department of Radiation Oncology\\
       Mayo Clinic, Rochester, MN, USA 55905
       \AND
       \Name{Andrew Y.K. Foong}
       \Email{Foong.Andrew@mayo.edu}\\ 
       \addr Department of Radiation Oncology\\
       Mayo Clinic, Rochester, MN, USA 55905} 

\begin{document}
\maketitle

\begin{abstract}
Survival analysis is an established framework for analyzing time-to-event data, yet many clinical machine learning studies still binarize the outcome before model training. This practice excludes censored patients, collapses temporal information into a single threshold, and can affect which features are selected as prognostically relevant. We examine the cost of this binarization in the context of Bayesian network (BN) feature selection, using two recent publications as case studies: one that applies BN-based feature selection to a head-and-neck cancer cohort and a second surgical cohort study that, while not BN-based, likewise binarizes its survival endpoint. We replace the binary scoring function with the Cox partial log-likelihood for feature-to-outcome edges, a modification we call the Survival-Aware Bayesian network, and recover prognostic features that binarization misses. Our ablation experiment confirms that the improvement is driven by the time-to-event scoring formulation rather than by retaining more patients. The results generalize across five endpoint-cohort combinations in head-and-neck cancer and extend to three further cancer types (breast, colorectal, and kidney). We propose that clinical studies with survival outcomes should use time-to-event methods by default, as binarization discards the prognostic signal retained by survival analysis.
\end{abstract}

\section{Introduction}
\label{sec:intro}

Survival analysis has been the standard framework for time-to-event outcomes, with well-understood methods for handling censoring, estimating hazard functions, and comparing treatment effects \citep{david2012survival}.
However, a large fraction of clinical machine learning studies still convert time-to-event outcomes into binary targets before predictive modeling. The practice is especially prevalent in Bayesian network (BN) modeling of cancer outcomes \citep{gevaert2006predicting, rabinowicz2017prognostic, shah2026bayesian}. For instance, in head-and-neck cancer (HNC), \citet{kotevski2023machine} predicted 2-year HNC survival as a binary endpoint across multiple institutions. \citet{kazmierski2023multi} chose binary 2-year survival for multi-institutional HNC modeling, noting that this formulation ``is commonly used in the literature." \\[\baselineskip]
Binarization, however, is not a neutral modeling choice. A time-to-event outcome, when collapsed into a fixed-horizon indicator, discards information about when events occur, inconsistently censors patients depending on follow-up length, and limits the use of hazard ratios and time-dependent risk estimates. These costs are well understood in the survival analysis literature. Still, they are rarely discussed in BN-based prognostic studies, where the graph structure and Markov blanket are otherwise leveraged for interpretable feature selection \citep{friedman1997bayesian, pearl2009causality, lucas2004bayesian, kyrimi2021comprehensive, mclachlan2020bayesian, tsamardinos2003markov,aliferis2010local}. \\[\baselineskip]
In this work, we quantify the cost of endpoint binarization using two head-and-neck cancer cohorts as case studies: one based on a well-established public dataset and the other drawn from a newly published cohort. The first case study is a Bayesian network prognostic model on a radiation cohort~\citep{shah2026bayesian} that binarizes survival for BN structure learning; the second is a prognostic analysis on a surgical cohort~\citep{dorrich2025multimodal} that does not use BNs but adopts the same endpoint binarization. For both studies, we show that binarizing the survival endpoint results in a measurable loss of prognostic information.
We demonstrate how survival analysis can be performed without sacrificing the interpretability advantages that motivated the use of BNs in the first place. We refer to our approach as Survival-Aware Bayesian network (SA-BN) and evaluate it on these two HNC cohorts, which differ in treatment modality (radiation, $n{=}2{,}994$; surgical, $n{=}763$), and ask three questions: (1)~Which features does binary scoring miss and why? (2)~Is the improvement driven by the Cox scoring formulation or by retaining more patients? (3)~Do the survival-selected features degrade binary classification performance?

\subsection*{Generalizable Insights about Machine Learning in the Context of Healthcare}

This work provides three insights relevant to studies that binarize a survival outcome: 

\begin{enumerate}[label=\roman*),leftmargin=*,nosep]
  \item \textbf{Endpoint binarization hides features with graded survival effects.} Fixed-horizon thresholds (e.g., 2-year survival) may not detect prognostic variables whose influence is continuous rather than concentrated at a cutpoint. In our case studies, switching from a binary to a survival-aware scoring function recovers established HNC prognostic factors that the binarization step misses, including smoking pack-years, overall stage, hemoglobin, neutrophil-to-lymphocyte ratio (NLR), positive lymph node count, and age. Any study that binarizes a survival outcome should verify that its selected features are not a consequence of this formulation.

  \item \textbf{The scoring function determines which features are considered prognostic.} A common response to censoring-induced exclusion is to obtain more data. In the radiation cohort, running the survival-aware pipeline on the same 2{,}635 patients with sufficient ($\geq$2 years) survival information recovers 8 of 9 Markov blanket features; the additional 359 patients with insufficient survival information contribute only one. The endpoint-scoring formulation primarily determines which features are considered prognostic.

  \item \textbf{Survival-aware feature selection matches or improves binary performance while enabling effect estimation in units of time.} When features selected under survival-aware scoring are used for binary 2-year classification, performance improves over features selected under binary scoring. The same features also support interventional RMST, which expresses each factor in months of expected life after confounder adjustment. A binary-proxy model cannot produce these quantities for any feature, including the ones it selects.
\end{enumerate}

\section{Related Work}
\label{sec:related}

\textbf{Binarization of survival endpoints.} The practice of converting time-to-event outcomes into binary classification targets is widespread in clinical ML.
In head-and-neck cancer alone, \citet{shah2026bayesian} used 2-year survival for BN-based feature selection on the RADCURE cohort \citep{welch2024radcure}. \citet{diamant2019deep} trained deep learning models on binary 2-year overall survival, and \citet{starke2023longitudinal} adopted the same endpoint for multi-institutional radiomics modeling. We hypothesize that in each case, the binary endpoint method was chosen for methodological convenience rather than clinical necessity. The clinical question concerns the time to death, which a fixed threshold only partially summarizes. No prior work has systematically quantified the cost of this binarization in terms of feature-selection quality or tested whether the lost features can be recovered by switching to a time-to-event scoring function.\\[\baselineskip]
\textbf{Bayesian network Markov blankets and the target specification problem.}
The Markov blanket (MB) of a BN is the minimal set of variables that makes the target conditionally independent of all other variables, with theoretical optimality guarantees under faithfulness~\citep{tsamardinos2003markov,aliferis2010local}. These guarantees are conditional on the target being correctly specified. We argue that when the target is a binary proxy for a continuous survival outcome, the MB is optimal for the proxy, not for survival itself. A separate line of work~\citep{scutari2019learns} has shown that no single BN structure-learning algorithm dominates across datasets and recommends ensemble approaches. Our method addresses both issues: we replace the binary target with a survival-aware scoring function and require consensus across three structurally different BN learning algorithms (Appendix~\ref{app:algo_detail}). \\[\baselineskip]
\textbf{Survival-native feature selection.}
Feature selection methods such as LASSO-penalized Cox regression ~\citep{tibshirani1997lasso},
random-survival-forest variable importance ~\citep{ishwaran2008rsf}, and Cox stepwise selection use time-to-event information without binarizing. Each returns a ranked or penalized feature list, leaving the conditional independence structure among features implicit. SA-BN returns a graph and its Markov blanket, a structure that the ranked selectors do not produce. This blanket serves as the adjustment set for G-computation \citep{keil2014parametric} to estimate the interventional RMST for each prognostic feature.

\section{Methods}
\label{sec:methods}

\label{sec:pipeline}

Our pipeline has four phases:
(1)~SA-BN structure learning on all patients (including censored) to discover the survival-aware Markov blanket (SA-MB). As a baseline, we compare against the binary-proxy approach of \citet{shah2026bayesian}, which learns BN structure on a discretized ``SVy2" target (survived $\geq$2~years).
(2)~Survival modeling with Cox~PH, RSF and Gradient Boosting Survival (GBS) models on the selected features \citep{cox1972regression, ishwaran2008rsf, hothorn2006survival}.
(3)~Causal inference using interventional RMST and hazard ratios to validate that the selected features have interpretable graded effects.
(4)~Ablation experiments to disentangle the contributions of the Cox scoring formulation from the inclusion of additional censored patients and to verify that survival-selected features do not degrade binary classification.

\subsection{Survival-Aware BN Structure Learning (SA-BN)}
\label{sec:sabn_methods}

Standard BN structure learning scores node families using a classification-based scoring function (e.g., BDeu \citep{heckerman1995learning} and K2 \citep{cooper1992bayesian}) that requires a discrete target.\footnote{These are sometimes called multinomial scores because they model the target's conditional distribution as a multinomial over discrete categories.}
SA-BN decouples feature--feature edges from feature--target edges, enabling the target node to represent time-to-event information. Feature--target edges are scored by the Cox partial log-likelihood and added by a likelihood-ratio test (LRT, $p < 0.05$), so censored patients contribute to structure learning through their time and event indicators rather than being excluded by a fixed-horizon cutpoint. All scoring equations, algorithm hyperparameters, and cycle-breaking rules are provided in Appendix~\ref{app:algo_detail}.

\subsection{Markov Blanket Discovery and Downstream Modeling}
\label{sec:smb_methods}
We identify the survival Markov blanket by consensus ($\geq$2/3) across three methods: DAG-derived parents, children, and co-parents of the survival target; Cox stepwise selection; and univariate C-index top-$k$. For comparison, we also extract the binary-proxy Markov blanket, following \citet{shah2026bayesian}, for the SVy2 target. Three survival models (Cox PH, Random Survival Forest, Gradient-Boosted Survival) are then trained on the selected features and evaluated with concordance index~\citep{harrell1996multivariable} (1{,}000-bootstrap 95\% CIs). Causal effects are estimated via interventional RMST by time-to-event G-computation \citep{keil2014parametric,hernan2020causal}, which fits a Cox model with the remaining Markov blanket parents as confounders, applies the do-operation to the feature of interest, and integrates the resulting individual survival curves to 5 years. Algorithm details, scoring thresholds, model hyperparameters, evaluation formulas, and G-computation derivations are in Appendices~\ref{app:algo_detail},~\ref{app:bn_figs}, and~\ref{app:gcomp}. Per-feature hazard-ratio $p$-values from the DAG-informed Cox models are corrected for multiple comparisons within each cohort using the Benjamini--Hochberg false-discovery-rate (FDR) procedure \citep{benjamini1995controlling}, and the proportional-hazards assumption is assessed by Schoenfeld-residual tests \citep{grambsch1994proportional, schoenfeld1982partial} with the same FDR control (Appendix~\ref{app:ph}).

\section{Cohort}
\label{sec:cohort}
\subsection{Cohorts and features }
The two HNC cohorts used in this study are summarized in Table~\ref{tab:cohorts}.\\
\\
\textbf{Cohort~1: Radiation:}
2,994 patients with HNC treated with definitive radiation at a single institution (2005--2017)~\citep{welch2024radcure}.
For binary BN learning, 2,635 patients are evaluable for 2-year survival (88\%), for survival modeling, all 2,994 are retained.
Temporal split at year 2015, training $n{=}2{,}174$, test $n{=}820$.
Event rate: 32.7\% (978 deaths). Thirteen features, including demographics (age, sex), tumor characteristics (T~stage, N~stage, primary tumor site, Gross Tumor Volume, Histological Subtype), patient factors (Eastern Cooperative Oncology Group Performance Status (ECOG PS), smoking pack-years, Human Papillomavirus status), treatment (modality), overall stage, and a T$\times$N interaction. \\
\\
\textbf{Cohort~2: Surgical:}
763 patients with HNC treated with primary surgery at a single institution (2005--2019) \citep{dorrich2025multimodal}. For binary BN learning, 620 are evaluable for 2-year survival (81\%) and for survival modeling, all 763 are retained.
Temporal split at year 2016, training $n{=}565$, test $n{=}198$.
Event rate: 27.9\% (213 deaths). Fifteen features consisting of primary tumor site, pT, pN, grade, HPV, sex, smoking, adjuvant radiotherapy, age, infiltration depth, positive lymph node count, blood hemoglobin, NLR, invasion burden composite, and T$\times$N interaction.
Six additional variables available in the raw dataset (resection status, resected lymph node count, resection margin, leukocytes, platelets, platelet-to-lymphocyte ratio (PLR)) were excluded before structure learning due to near-constant values (resection status: 90\% R0) or high collinearity with retained features (leukocytes, platelets, and PLR with NLR; resected count with positive count).

\begin{table}[t]
\centering
\caption{Cohort characteristics. Survival modeling retains all patients; binary BN learning uses evaluable patients only.}
\label{tab:cohorts}
\begin{tabular}{@{}lll@{}}
\toprule
& \textbf{Radiation} & \textbf{Surgical} \\
\midrule
Treatment & Definitive RT & Surgery primary \\
Total patients & 2,994 & 763 \\
Binary-evaluable & 2,635 & 620 \\
Features (BN) & 13 & 15 \\
Temporal split & 2015 & 2016 \\
Train / test (surv.) & 2,174 / 820 & 565 / 198 \\
Event rate & 32.7\% & 27.9\% \\
Has HPV & Yes & Yes \\
Has invasion markers & No & Yes \\
\bottomrule
\end{tabular}
\end{table}

\section{Results}
\label{sec:results}

\subsection{SA-BN Recovers Features That Binary Scoring Misses}
\label{sec:sabn_results}

\begin{figure}[ht]
\centering
\includegraphics[width=0.7\linewidth]{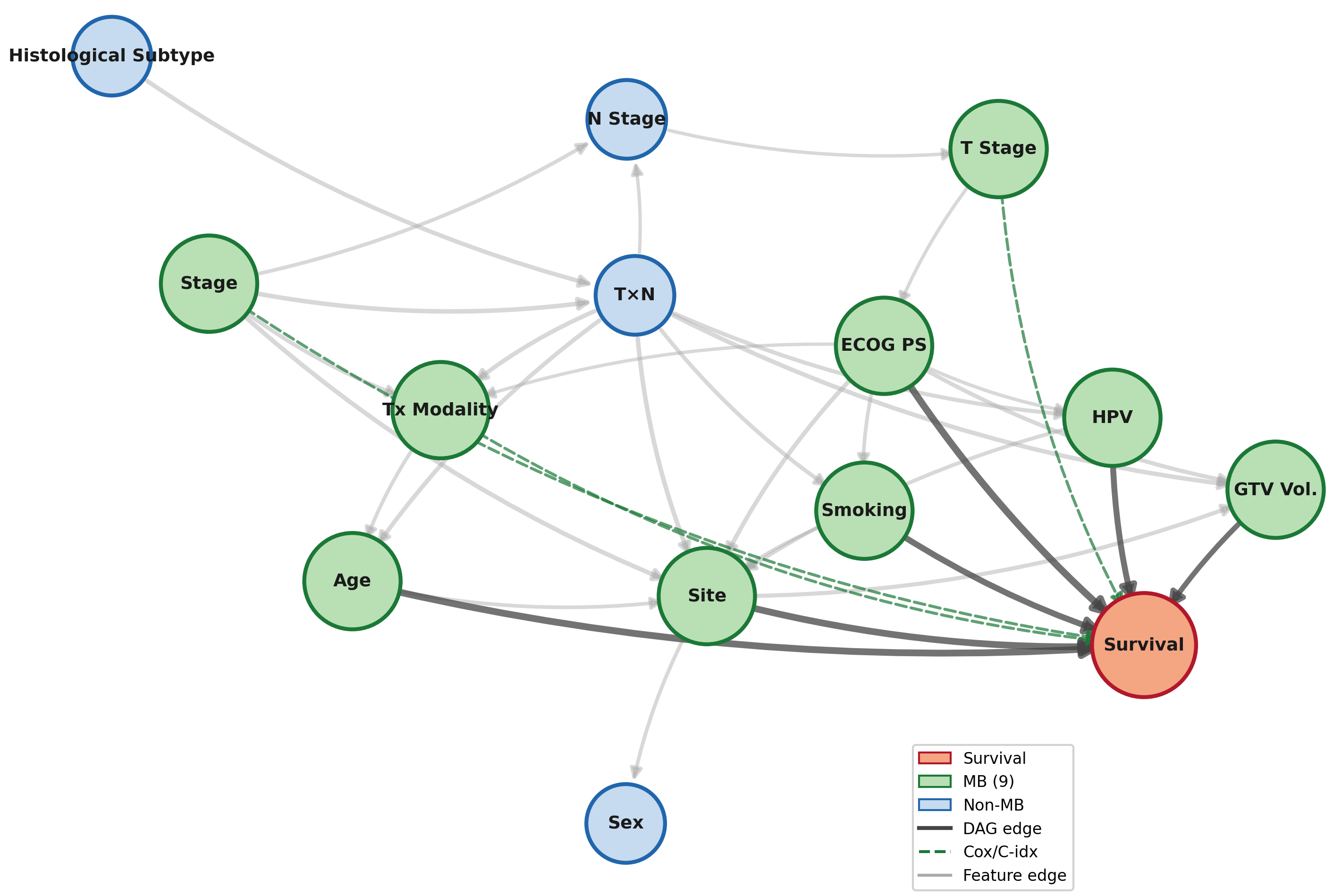}
\caption{SA-BN consensus structure. Radiation cohort (14 nodes, 28 edges, 9-feature Markov blanket). Orange = survival target, green = consensus survival Markov blanket, blue = non-Markov blanket. Dark arrows = DAG-derived edges to survival; dashed green = Cox/C-index selected; gray = feature--feature edges.}
\label{fig:bn_rad}
\end{figure}

\begin{figure}[ht]
\centering
\includegraphics[width=0.7\linewidth]{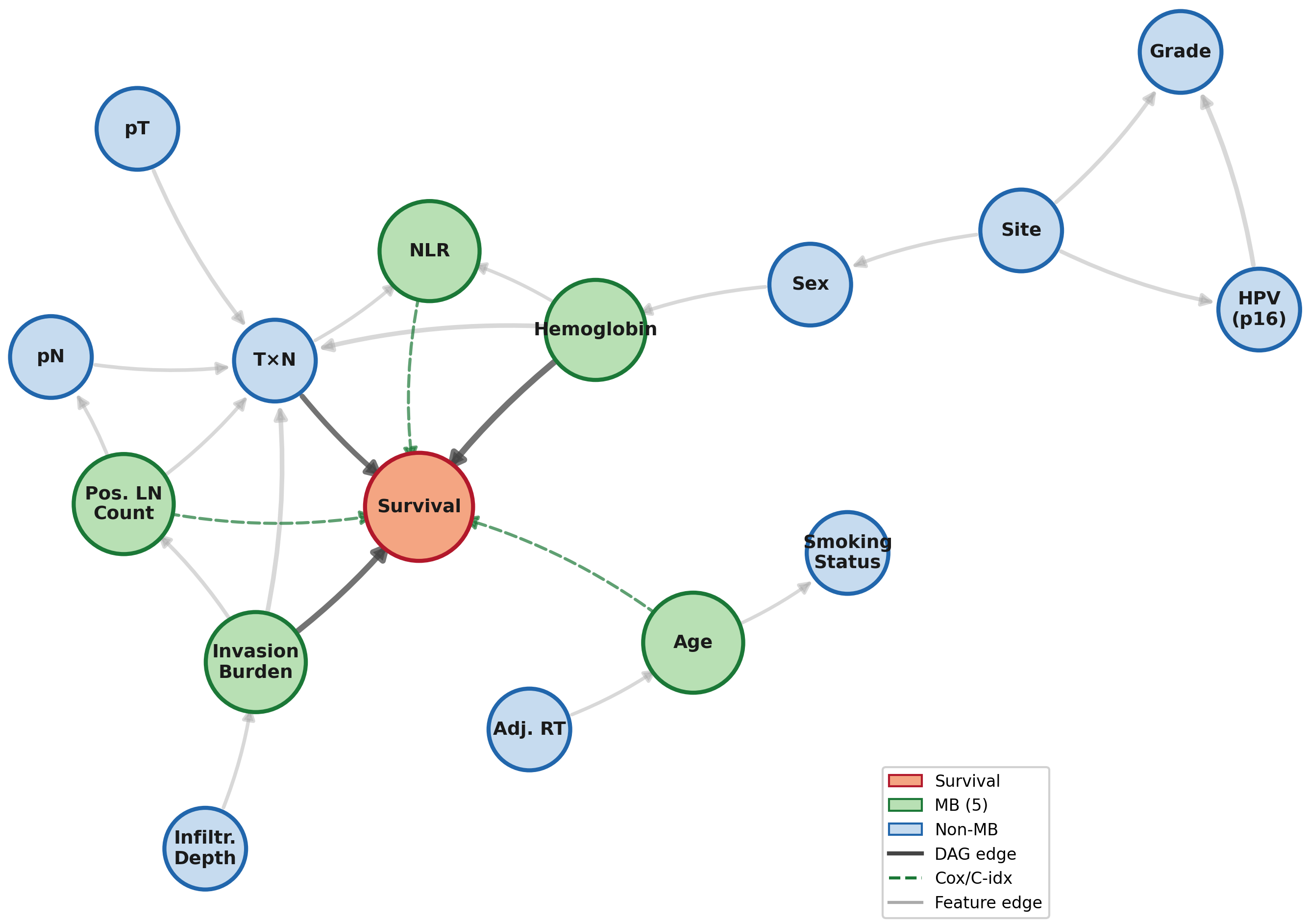}
\caption{SA-BN consensus structure. Surgical cohort (16 nodes, 25 edges, 5-feature Markov blanket). Orange = survival target, green = consensus survival Markov blanket, blue = non-Markov blanket. Dark arrows = DAG-derived edges to survival; dashed green = Cox/C-index selected; gray = feature--feature edges.}
\label{fig:bn_han}
\end{figure}

\begin{figure}[ht]
\centering
\subfigure[Radiation Cohort: 7/9 Markov blanket at 100\%; stage (68\%), T (66\%) moderate.]{\includegraphics[height=5.2cm]{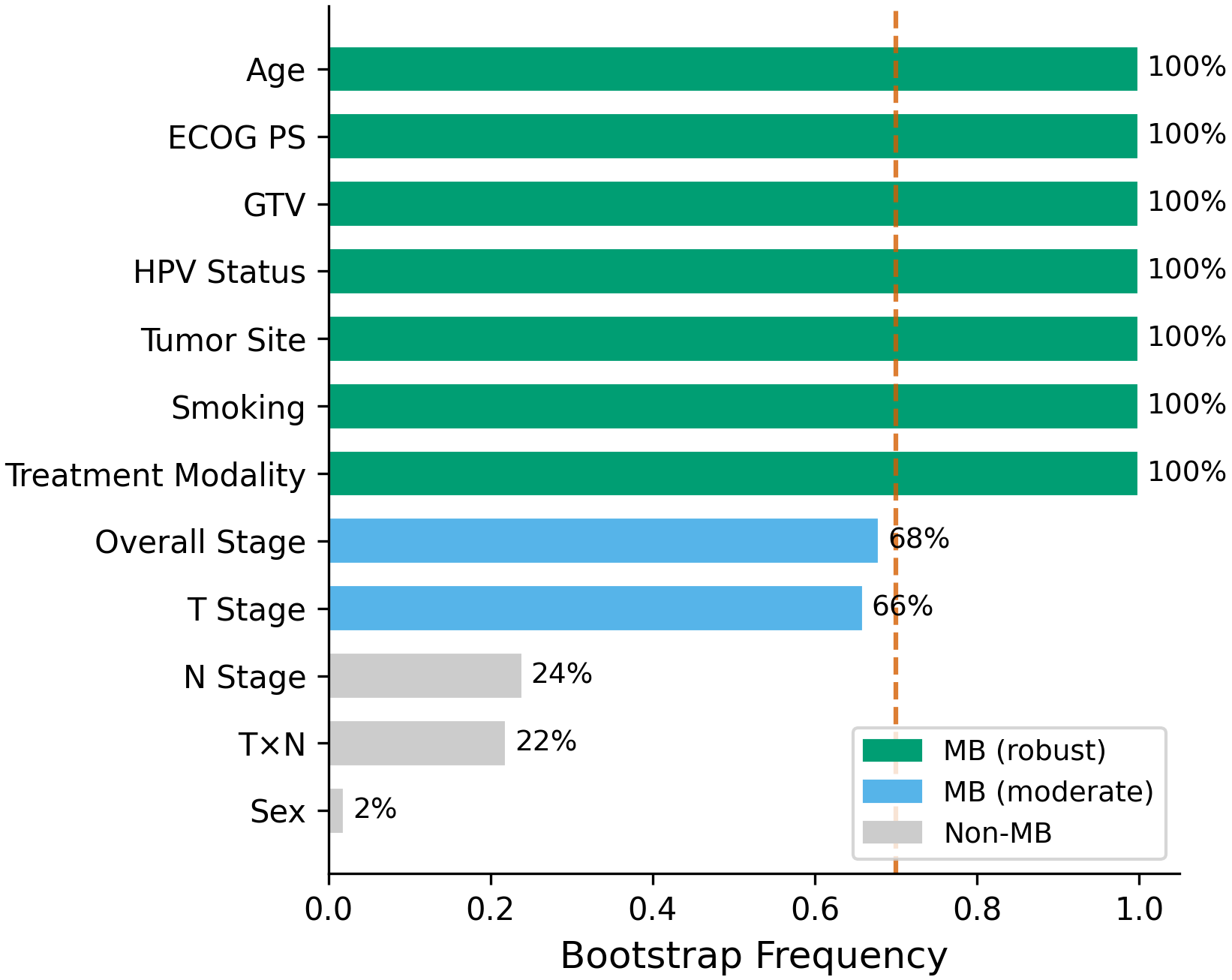}}
\subfigure[Surgical Cohort: all 5 Markov blanket members above 78\%.]{\includegraphics[height=5.2cm]{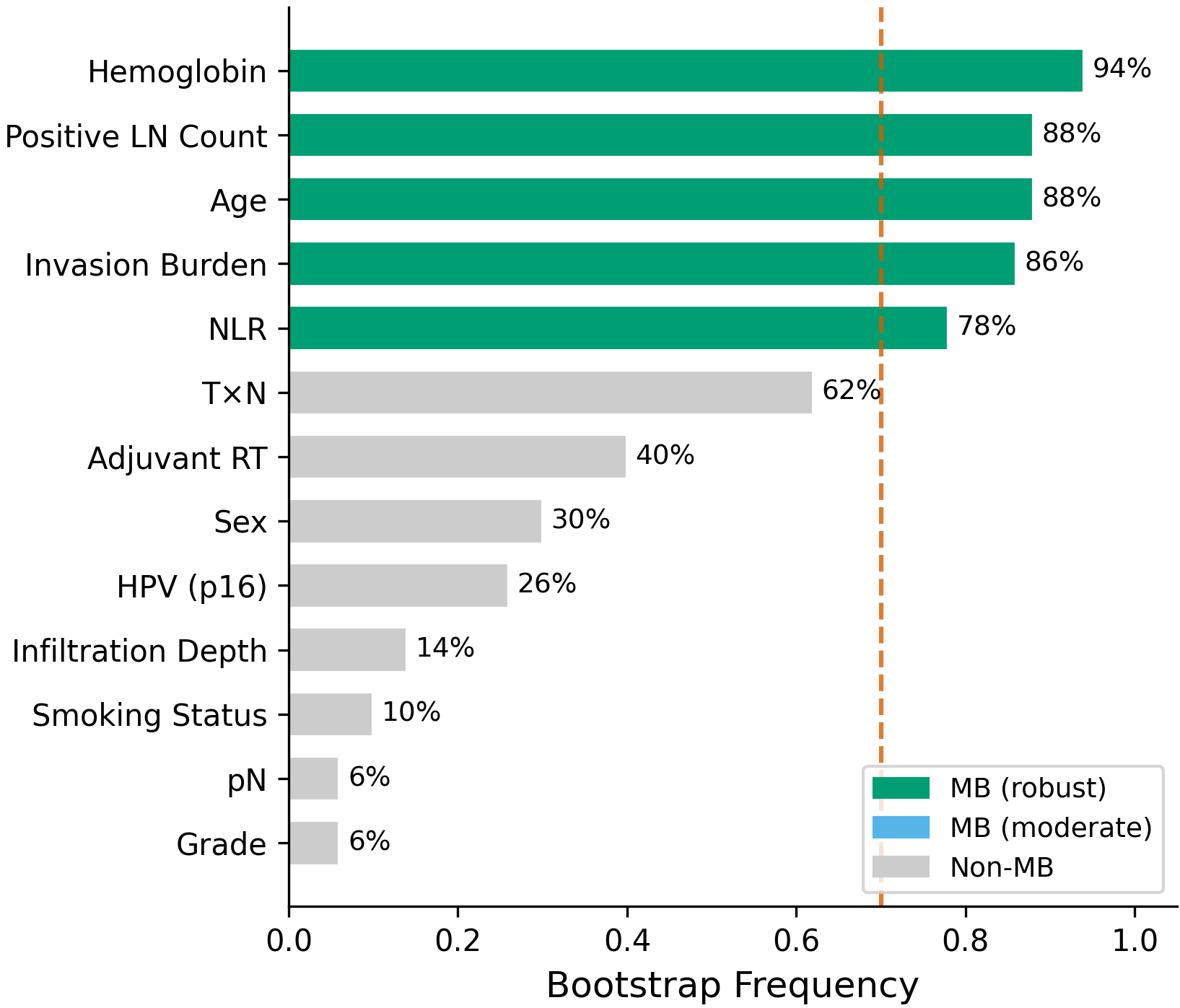}}
\caption{SA-BN Markov blanket stability across 50 bootstrap resamples. Dashed line = 70\% threshold.}
\label{fig:mb_stability}
\end{figure}

The binary-proxy and survival-aware Markov blankets are compared and summarized in Table~\ref{tab:sabn_comparison}.
In the radiation cohort, binary scoring excludes 359 patients (12\%) and discovers an 8-feature Markov blanket.
SA-BN retains all patients and discovers a 9-feature survival Markov blanket (Figure~\ref{fig:bn_rad}). The features unique to SA-BN that no binary implementation recovers are smoking pack-years and overall HNC stage, both with continuous, graded-effect relationships that binarization collapses~\citep{ang2010human,ma2022association}. SA-BN also drops the T$\times$N interaction, whose signal is subsumed by the T stage and the overall stage separately. In the surgical cohort, SA-BN replaces the binary Markov blanket's coarse categorical variables (grouped pN and infiltration depth) with continuous alternatives, such as the actual count of positive lymph nodes, blood hemoglobin, and NLR (Figure~\ref{fig:bn_han}). The SA-BN Markov blanket improves test-set concordance index by +0.031 (radiation) and +0.078 (surgical) while retaining $>$95\% of all-feature performance in both cohorts. Seven of nine radiation features and all five surgical features exceed the $70\%$ robustness threshold across 50 bootstrap resamples (Figure~\ref{fig:mb_stability}), confirming that the feature selection is stable. Per-feature selection frequencies, hazard ratios, BH-FDR $q$-values, and directed-edge stabilities are reported in Table~\ref{tab:quant_mb} (Appendix~\ref{app:quant_mb}); all features survive multiple-comparison correction at $q<0.05$ except radiation T~stage.

\begin{table}[t]
\centering
\footnotesize
\caption{All candidate features and their Markov blanket membership. $\bullet$ = selected, -- = not selected. SA-BN test-set C-index: radiation 0.801 [.76--.84], surgical 0.726 [.64--.81]; binary: 0.770 [.73--.81], 0.648 [.58--.75].}
\label{tab:sabn_comparison}
\begin{tabular}{@{}lcc|lcc@{}}
\toprule
\multicolumn{3}{c|}{\textbf{Radiation Cohort (13 features)}} & \multicolumn{3}{c}{\textbf{Surgical Cohort (15 features)}} \\
\textbf{Feature} & \textbf{Binary BN} & \textbf{SA-BN} & \textbf{Feature} & \textbf{Binary BN} & \textbf{SA-BN} \\
\midrule
Age & $\bullet$ & $\bullet$ & Tumor Site & -- & -- \\
ECOG PS & $\bullet$ & $\bullet$ & pT (grouped) & -- & -- \\
GTV & $\bullet$ & $\bullet$ & pN (grouped) & $\bullet$ & -- \\
HPV status & $\bullet$ & $\bullet$ & Grade & -- & -- \\
Tumor Site & $\bullet$ & $\bullet$ & HPV (p16) & -- & -- \\
T stage & $\bullet$ & $\bullet$ & Sex & -- & -- \\
Treatment modality & $\bullet$ & $\bullet$ & Smoking & -- & -- \\
T$\times$N & $\bullet$ & -- & Adjuvant\ radiotherapy & -- & -- \\
Sex & -- & -- & Age & -- & $\bullet$ \\
N stage & -- & -- & Infiltration\ depth & $\bullet$ & -- \\
Histological Subtype & -- & -- & Positive\ Lymph Node count & -- & $\bullet$ \\
Smoking (pack-yr) & -- & $\bullet$ & Hemoglobin & -- & $\bullet$ \\
Overall stage & -- & $\bullet$ & NLR & -- & $\bullet$ \\
 & & & Invasion burden & $\bullet$ & $\bullet$ \\
 & & & T$\times$N & -- & -- \\
\bottomrule
\end{tabular}
\end{table}

\subsection{Cox scoring recovers the Markov blanket on the binary-evaluable patients}
\label{sec:ablation}

Two factors could explain the difference in the Markov blanket: Cox scoring (formulation) and additional censored patients (data). In the radiation cohort, running SA-BN on the 2,635 patients (excluding 359 censored before 2 years) recovers 8 of 9 survival MB features (Table~\ref{tab:ablation}), and only the overall stage is dropped. In the surgical cohort, running SA-BN on the 620 patients (excluding 143 censored before 2 years) recovers all 5 of 5 features and adds T$\times$N. In both cohorts, the Cox scoring formulation is the primary driver, and the extra censored patients contribute at most one feature. 

\begin{table}[ht]
\centering
\footnotesize
\caption{Ablation: all candidate features and their MB membership across three conditions. $\bullet$ = selected, -- = not selected. Binary uses evaluable patients only (2,635/620); SA-BN (eval.) uses the same patients with Cox scoring; SA-BN (all) adds censored patients.}
\label{tab:ablation}
\begin{tabular}{@{}lccc|lccc@{}}
\toprule
\multicolumn{4}{c|}{\textbf{Radiation Cohort (13 features)}} & \multicolumn{4}{c}{\textbf{Surgical Cohort (15 features)}} \\
\textbf{Feature} & \rotatebox{70}{\textbf{Binary}} & \rotatebox{70}{\textbf{SA-BN (ev.)}} & \rotatebox{70}{\textbf{SA-BN (all)}} & \textbf{Feature} & \rotatebox{70}{\textbf{Binary}} & \rotatebox{70}{\textbf{SA-BN (ev.)}} & \rotatebox{70}{\textbf{SA-BN (all)}} \\
\midrule
Age            & $\bullet$ & $\bullet$ & $\bullet$ & Tumor Site            & --        & --        & --        \\
ECOG PS        & $\bullet$ & $\bullet$ & $\bullet$ & pT (grouped)    & --        & --        & --        \\
GTV     & $\bullet$ & $\bullet$ & $\bullet$ & pN (grouped)    & $\bullet$ & --        & --        \\
HPV status     & $\bullet$ & $\bullet$ & $\bullet$ & Grade           & --        & --        & --        \\
Tumor Site           & $\bullet$ & $\bullet$ & $\bullet$ & HPV (p16)       & --        & --        & --        \\
T stage        & $\bullet$ & $\bullet$ & $\bullet$ & Sex             & --        & --        & --        \\
Treatment modality    & $\bullet$ & $\bullet$ & $\bullet$ & Smoking         & --        & --        & --        \\
T$\times$N     & $\bullet$ & --        & --        & Adjuvant\ RT        & --        & --        & --        \\
Sex            & --        & --        & --        & Age             & --        & $\bullet$ & $\bullet$ \\
N stage        & --        & --        & --        & Infiltration\ depth & $\bullet$ & --        & --        \\
Histological Subtype      & --        & --        & --        & Positive \ LN count  & --        & $\bullet$ & $\bullet$ \\
Smoking (pk-yr)& --        & $\bullet$ & $\bullet$ & Hemoglobin      & --        & $\bullet$ & $\bullet$ \\
Overall stage  & --        & --        & $\bullet$ & NLR             & --        & $\bullet$ & $\bullet$ \\
               &           &           &           & Invasion\ burden    & $\bullet$ & $\bullet$ & $\bullet$ \\
               &           &           &           & T$\times$N      & --        & $\bullet$ & --        \\
\midrule
$|$MB$|$       & 8         & 8         & 9         & $|$MB$|$        & 3         & 6         & 5         \\
\bottomrule
\end{tabular}
\end{table}

\subsection{Switching to survival MB features is also beneficial for binary classification}
\label{sec:binary_validation}

We then ask whether features selected for survival prediction remain useful for binary 2-year classification.
Table~\ref{tab:binary_cls} compares the two MBs on binary prediction for both cohorts. In the radiation cohort, the SA-BN MB improves binary classification on all metrics: AUC-ROC from $0.797$ to $0.826$ (+0.029), AUC-PR from $0.943$ to $0.950$, and MCC from $0.369$ to $0.407$.
The SA-BN features (which include smoking and overall stage) provide information that helps discriminate the 2-year thresholds and predict survival time. In the surgical cohort, the SA-BN MB also improves binary classification: AUC-ROC from $0.740$ to $0.807$ and MCC from $0.266$ to $0.487$. The test set here is small (35 events), so the bootstrap confidence intervals overlap, and this improvement is directional. Across both cohorts, the survival-aware feature set improves performance on the binary 2-year task while also supporting survival curves and interventional RMST, which a binarized endpoint cannot provide. The survival-modeling advantage is model-dependent: in the surgical cohort, linear Cox and LASSO-Cox slightly favor the binary feature set on discrimination (Appendix~\ref{app:model_perf}), and the survival-aware gain comes from RSF and the integrated Brier score.

\begin{table}[t]
\centering
\footnotesize
\setlength{\tabcolsep}{4pt}
\caption{Binary 2-year classification on evaluable patients. Mean $\pm$ std, 200 bootstrap resamples. Bold: best per cohort per model. LR: Logistic Regression; RF: Random Forest}
\label{tab:binary_cls}
\begin{tabular}{@{}lllcccc@{}}
\toprule
\textbf{Cohort} & \textbf{Features} & \textbf{Model} & \textbf{AUC-ROC} & \textbf{AUC-PR} & \textbf{F1} & \textbf{MCC} \\
\midrule
\multirow{4}{*}{Radiation}
& Binary MB (8) & LR & $.797 \pm .024$ & $.943 \pm .011$ & $.815 \pm .016$ & $.369 \pm .041$ \\
& Binary MB (8) & RF & $.785 \pm .026$ & $.938 \pm .012$ & $.828 \pm .016$ & $.329 \pm .043$ \\
& SA-BN MB (9) & LR & $\mathbf{.826 \pm .023}$ & $\mathbf{.950 \pm .011}$ & $\mathbf{.836 \pm .015}$ & $\mathbf{.407 \pm .043}$ \\
& SA-BN MB (9) & RF & $\mathbf{.820 \pm .024}$ & $\mathbf{.948 \pm .011}$ & $\mathbf{.849 \pm .014}$ & $\mathbf{.404 \pm .045}$ \\
\midrule
\multirow{4}{*}{Surgical}
& Binary MB (3) & LR & $.697 \pm .061$ & $.903 \pm .029$ & $.843 \pm .024$ & $.231 \pm .094$ \\
& Binary MB (3) & RF & $.740 \pm .053$ & $.925 \pm .023$ & $.860 \pm .023$ & $.266 \pm .097$ \\
& SA-BN MB (5) & LR & $.737 \pm .058$ & $.933 \pm .021$ & $.872 \pm .022$ & $.239 \pm .097$ \\
& SA-BN MB (5) & RF & $\mathbf{.807 \pm .052}$ & $\mathbf{.950 \pm .019}$ & $\mathbf{.935 \pm .015}$ & $\mathbf{.487 \pm .102}$ \\
\bottomrule
\end{tabular}
\end{table}

\subsection{Feature-Level Interventional RMST}
\label{sec:rmst}
Binary BNs can report P(SVy2 = 1 $|$ do(X = x)) via do-calculus~\citep{pearl2009causality, shah2026bayesian}, but this collapses survival beyond 2~years into a single probability.
The survival-endpoint analog, interventional RMST via G-computation, preserves the full temporal structure and expresses effects in months of life (Figures~\ref{fig:rmst_levels_rad} and~\ref{fig:rmst_levels_han}). In the radiation cohort, moving from ECOG~PS~0 to PS~3+ is associated with a loss of 8~months of interventional RMST after adjustment for confounders, consistent with the established prognostic role of performance status in HNC~\citep{chalker2022performance}. HPV-positive patients gain 8~months of interventional RMST relative to HPV-negative patients, in line with the RTOG~0129 finding that HPV is the strongest independent prognostic factor in oropharyngeal cancer~\citep{ang2010human}. In the surgical cohort, hemoglobin $\leq$12.1~g/dL costs 13.9~months of interventional RMST relative to 15.1--16.2~g/dL, reflecting the well-documented dose-dependent association between pretreatment anemia and survival in HNC~\citep{prosnitz2005pretreatment,ma2022defining}. Positive lymph node count $\geq$8 costs 12~months, mirroring the continuous mortality gradient reported for metastatic nodal burden in oral cavity cancer~\citep{roberts2016number,lee2019number}. NLR $\geq$4.4 costs 7~months, consistent with the meta-analytic HR of 1.69--1.84 for elevated NLR across HNC cohorts~\citep{yang2018prognostic,takenaka2018prognostic}. All SA-BN unique features (except T stage) retain monotone graded effect gradients after confounding adjustment.
DAG-informed hazard ratios confirm these patterns in standard clinical units of Hazard Ratios (HR) (ECOG PS HR 1.49--4.22; invasion burden HR 1.22--2.45; hemoglobin HR 0.85; Table~\ref{tab:hazard_ratios}, Appendix~\ref{app:causal_figs}).

\subsection{Generalization Across Survival Endpoints}
\label{sec:endpoints}

We repeat the SA-BN analysis on loco-regional recurrence (radiation cohort: 303 events, 10.1\%; 22.3\% excluded by binarization), PFS (surgical cohort: 297 events, 39\%) and recurrence (surgical cohort: 177 events, 23\%).
Table~\ref{tab:endpoints} shows that the binarization cost applies across all endpoint types. In the radiation cohort, the recurrence Markov blanket shares 4 of 9 OS features and drops OS-specific factors (HPV, smoking, staging). In the surgical cohort, 4 of 5 OS members reappear in PFS; positive LN count is the only feature shared across all three surgical endpoints. Continuous biomarkers (hemoglobin, NLR, positive LN count) consistently appear in the SA-BN Markov blanket and are consistently missed by other methods, regardless of endpoint.

\begin{table}[t]
\centering
\footnotesize
\caption{SA-BN Markov blanket membership across survival endpoints. $\bullet$ = selected, -- = not selected. Features appearing across multiple endpoints are more robust to endpoint choice.}
\label{tab:endpoints}
\begin{tabular}{@{}lcc|lcccc@{}}
\toprule
\multicolumn{3}{c|}{\textbf{Radiation cohort}} & \multicolumn{4}{c}{\textbf{Surgical cohort}} \\
\textbf{Feature} & \textbf{OS} & \textbf{Recurrence} & \textbf{Feature} & \textbf{OS} & \textbf{PFS} & \textbf{Recurrence} \\
\midrule
Age & $\bullet$ & -- & Age & $\bullet$ & $\bullet$ & -- \\
ECOG PS & $\bullet$ & $\bullet$ & Hemoglobin & $\bullet$ & $\bullet$ & -- \\
GTV & $\bullet$ & $\bullet$ & NLR & $\bullet$ & $\bullet$ & -- \\
HPV status & $\bullet$ & -- & Positive\ LN count & $\bullet$ & $\bullet$ & $\bullet$ \\
Tumor Site & $\bullet$ & $\bullet$ & Invasion burden & $\bullet$ & -- & $\bullet$ \\
T stage & $\bullet$ & -- & T$\times$N & -- & $\bullet$ & -- \\
Treatment modality & $\bullet$ & $\bullet$ & Smoking & -- & $\bullet$ & -- \\
Smoking (pack-yr) & $\bullet$ & -- & Adjuvant\ RT & -- & -- & $\bullet$ \\
Overall stage & $\bullet$ & -- & pT (grouped) & -- & -- & -- \\
T$\times$N & -- & $\bullet$ & pN (grouped) & -- & -- & -- \\
N stage & -- & -- & Grade & -- & -- & -- \\
Sex & -- & -- & HPV (p16) & -- & -- & -- \\
Histological Subtype & -- & -- & Sex & -- & -- & -- \\
 & & & Infiltration depth & -- & -- & -- \\
 & & & Tumor Site & -- & -- & -- \\
\midrule
$|$MB$|$ & 9 & 5 & $|$MB$|$ & 5 & 6 & 3 \\
Events & 978 & 303 & Events & 213 & 297 & 177 \\
\bottomrule
\end{tabular}
\end{table}

\subsection{Generalization Across Cancer Types}
\label{sec:tcga}

Both primary cohorts are single-institution and head-and-neck. To test whether the binarization cost is specific to that setting, we repeat the comparison on three multi-institutional cohorts from The Cancer Genome Atlas (TCGA) spanning different cancer types: breast (BRCA), colorectal (COAD) and kidney (KIRC), using the same 2-year binarization and the same SA-BN pipeline (Table~\ref{tab:tcga}). On all three, the survival-aware Markov blanket recovers prognostic variables the binary endpoint drops and leads on every metric. In kidney cancer, binarization collapses the blanket to overall stage alone, while SA-BN additionally recovers age, grade and nodal/metastatic stage; in breast cancer, the binary blanket latches onto treatment variables (adjuvant pharmacotherapy and radiotherapy) while SA-BN selects biology (age, menopausal status). The magnitude of the gain tracks the fraction of patients that binarization excludes (BRCA and COAD drop $\sim$40\% and gain $+0.06$/$+0.05$ C-index; KIRC drops 17\% and gains $+0.03$), consistent with the mechanism described in Section~\ref{sec:ablation} and the boundary conditions in Section~\ref{sec:scope}.

\begin{table}[t]
\centering
\footnotesize
\caption{Generalization to three multi-institutional TCGA cohorts at the 2-year endpoint. Each cohort reports the binary-proxy and survival-aware Markov blankets, along with their test-set metrics (C-index and time-dependent AUC, higher is better; integrated Brier score, lower is better). ``\% drop'' is the fraction of patients excluded by binarization.}
\label{tab:tcga}
\setlength{\tabcolsep}{4pt}
\begin{tabular}{@{}llccc@{}}
\toprule
\textbf{Cohort (n, \% drop)} & \textbf{Feature set} & \textbf{C-index} & \textbf{tAUC} & \textbf{IBS$\downarrow$} \\
\midrule
\multirow{2}{*}{BRCA breast (1083, 41\%)}
 & Binary (5): adj.\ pharm/RT, N, PR, stage & 0.652 & 0.647 & 0.081 \\
 & SA-BN (5): age, menopause, N, PR, stage & \textbf{0.711} & \textbf{0.724} & \textbf{0.077} \\
\midrule
\multirow{2}{*}{COAD colorectal (605, 40\%)}
 & Binary (2): M stage, pos.\ LN & 0.649 & 0.668 & 0.110 \\
 & SA-BN (2): age, M stage & \textbf{0.702} & \textbf{0.738} & \textbf{0.105} \\
\midrule
\multirow{2}{*}{KIRC kidney (535, 17\%)}
 & Binary (1): overall stage & 0.741 & 0.785 & 0.143 \\
 & SA-BN (5): age, grade, N, M, stage & \textbf{0.767} & \textbf{0.809} & \textbf{0.142} \\
\bottomrule
\end{tabular}
\end{table}


\section{Discussion}
\label{sec:discussion}

\subsection{Binarization Distorts the Markov Blanket}
Binarization of time-to-event information fails to identify features that oncologists already consider established prognostic factors for HNC. The ablation in Section~\ref{sec:ablation} shows this failure is caused by the scoring function itself, not by sample size. The mechanism is twofold: binarization excludes censored patients (thereby biasing the training set) and collapses continuous, graded effect relationships into a single threshold (making graded effects invisible). We illustrate three examples of how this plays out in practice. i) \emph{Smoking (pack-years)} in the radiation cohort: Smoking is an established prognostic factor in HNC, with a continuous graded effect relationship: heavier smokers have progressively worse survival. Neither \citet{shah2026bayesian} nor our binary baseline selects smoking, because binarizing at 2~years collapses the graded relationship between cumulative tobacco exposure and survival time into a single threshold. SA-BN recovers smoking in 100\% of bootstrap resamples because the Cox partial log-likelihood preserves the continuous gradient. ii) \emph{Hemoglobin} in the surgical cohort: Anemia is a known poor prognostic factor in HNC \citep{hoff2012importance} and its effect is graded: each 1 g/dL increase in hemoglobin multiplies the hazard by 0.85 (Table \ref{tab:hazard_ratios}). Binary discretization collapses this continuous graded effect into a single threshold, rendering the gradient invisible. iii) \emph{NLR} in the surgical cohort: Elevated neutrophils relative to lymphocytes is a blood-based marker of systemic inflammation, indicating a pro-tumor immune environment and impaired anti-tumor surveillance. NLR has a continuous, monotone effect on survival (interventional RMST: 4.30~yr at NLR $\leq$2.6 vs.\ 3.76~yr at NLR $\geq$4.4, a 7-month gap after confounder adjustment). Binary scoring misses NLR entirely because this gradient does not concentrate at any single time threshold. SA-BN recovers it at 78\% bootstrap stability. All three examples share the same mechanism; the feature has a continuous, graded effect on survival time that binary scoring cannot detect because the signal is spread across the entire survival curve rather than concentrated at the 2-year threshold.


\subsection{Clinical Implications}

The immediate downstream benefit of survival-aware scoring is that causal effects can be reported in units of time rather than probability. Applying G-computation to the SA-BN Markov blanket yields an interventional RMST for each feature level, which translates directly into months of expected life gained or lost after adjustment for confounders. Hemoglobin, missed by the binary-proxy BN in the surgical cohort, shows a monotonic dose--response in interventional RMST. NLR, also missed by binary scoring, shows the same pattern across its range. These gradients are precisely the signal that binarization erases; neither feature has a sharp effect at the 2-year cutpoint, but both substantially shift survival time across their full ranges. A binary-proxy BN cannot produce these quantities, even for the features it selects. When do-calculus is applied to the SVy2 target, it yields $P(\text{SVy2}{=}1 \mid \mathrm{do}(X{=}x))$, which collapses the entire post-2-year survival experience into a single probability and offers no way to distinguish a feature that shortens life by 2~months from one that shortens it by 12. 

\subsection{Integrating SA-BN into Clinical Workflows}

SA-BN is a drop-in change to an existing feature-selection step, and it fits clinical pipelines in three ways. First, the survival-aware Markov blanket is a compact, intake-level risk panel built from variables already recorded at first presentation (age, ECOG performance status, smoking history, blood hemoglobin, nodal count), and it can be scored without additional data collection. Second, the G-computation interventional RMST turns each blanket feature into a patient-facing quantity, e.g., months of expected life under a modifiable exposure, which is more directly actionable at the bedside than a shift in 2-year survival probability; pre-operative anemia correction (hemoglobin) is the clearest such example in the surgical cohort. Third, the blanket can serve purely as a feature-selection front-end ahead of an institution's existing Cox or RSF prognostic model, supplying a smaller, survival-justified set of covariates without changing the downstream model or its calibration. Each use replaces a binarized feature set with a survival-aware one, and the added computational cost is modest, dominated by a one-time offline stability analysis (Appendix~\ref{app:runtime}).

\subsection{Scope and Limitations of SA-BN}
\label{sec:scope}
The advantage of SA-BN is greatest when the dataset contains features with graded survival effects (e.g., hemoglobin, smoking pack-years) and substantial censoring. In the radiation cohort (33\% event rate, 12\% censored-excluded), SA-BN gains +0.031 C-index. In the surgical cohort (28\% event rate, 19\% censored-excluded), it increases by 0.078. The generalization across five endpoint-cohort combinations (Section~\ref{sec:endpoints}) suggests this is not endpoint-specific. If censoring is minimal or all features are binary (e.g., treatment received yes/no), SA-BN and binary BN will likely produce similar BN structures.
We also compare SA-BN against established survival-native selectors that avoid binarization: LASSO-Cox and RSF. At matched model size, a paired bootstrap
C-index difference test with BH-FDR \citep{benjamini1995controlling} does not distinguish SA-BN from a size-matched
LASSO-Cox or RSF variable-importance set on either cohort (all $q > 0.5$;
Appendix~\ref{app:parity}). The only significant difference favors a larger,
cross-validated 12-feature LASSO-Cox over the 6-feature radiation blanket
($q = 0.04$), where the estimator is fixed, and the feature count varies. SA-BN
matches these selectors on prediction and adds what they do not produce: a Markov
blanket that serves as the adjustment set for the interventional RMST analysis in
Section~\ref{sec:rmst}.

\paragraph{Limitations.}
(1)~The surgical test set contains only 35 events, so bootstrap CIs on the SA-BN vs.\ binary C-index comparison (0.726 vs.\ 0.648) overlap. The direction of the effect is consistent with the radiation cohort, but a larger surgical cohort is needed to confirm the magnitude.
(2)~SA-BN is a hybrid method: feature--feature edges use BDeu/K2 scoring while target--parent edges use Cox scoring. A fully survival-native scoring function for all edges would be more coherent but requires new theoretical development.

\section{Conclusion}
\label{sec:conclusion}

The binarization of survival outcomes for BN structure learning is a widespread practice in clinical machine learning; our experiments show that this practice incurs a measurable cost. In the two HNC cohorts, binary scoring misses established prognostic factors (smoking pack-years, overall stage, hemoglobin, NLR) because their effects are graded rather than threshold-based, producing unstable Markov blankets that vary across implementations. The fix requires updating how the graph scores the link between each feature and the survival outcome, while leaving the rest of the pipeline unchanged. A BN trained to predict how long patients survive, rather than whether they cross a 2-year threshold, recovers the established prognostic features and improves test-set performance. Causal analysis confirms that the recovered features have graded effects after adjusting for confounders. A controlled ablation shows that 8 of 9 features are recovered even on the same patients used by the binarized analysis, so the improvement comes from how the BN scores time-to-event survival information rather than from additional data. The survival-selected features also improve binary 2-year classification. The broader lesson is that the choice of endpoint formulation during BN structure learning, not just during model fitting, determines which features are considered prognostic. Clinical studies that binarize a survival outcome should verify that their Markov blanket is not an artifact of that binarization. \\

\paragraph{Data and Code availability.}
The radiation primary cohort is derived from the RADCURE dataset \citep{welch2024radcure} and is available for download from \url{https://www.cancerimagingarchive.net/collection/radcure/}.
The surgical cohort is derived from the HANCOCK dataset \citep{dorrich2025multimodal} and is available for download from \url{https://hancock.research.fau.eu/}.
Code is available at  \url{https://github.com/xinformatics/COBS}.


\bibliography{references}

@article{friedman1997bayesian,
  title={Bayesian network classifiers},
  author={Friedman, Nir and Geiger, Dan and Goldszmidt, Moises},
  journal={Machine learning},
  volume={29},
  number={2},
  pages={131--163},
  year={1997},
  publisher={Springer}
}

@book{pearl2009causality,
  title={Causality},
  author={Pearl, Judea},
  year={2009},
  publisher={Cambridge university press}
}

@article{shah2026bayesian,
  title={Clinically interpretable survival risk stratification in head and neck cancer using bayesian networks and markov blankets},
  author={Shah, Keyur D and Chamseddine, Ibrahim and Yuan, Xiaohan and Tian, Sibo and Qiu, Richard and Zhou, Jun and Dhabaan, Anees and Al-Hallaq, Hania and Yu, David S and Paganetti, Harald and others},
  journal={International Journal of Radiation Oncology* Biology* Physics},
  year={2025},
  publisher={Elsevier}
}

@article{gevaert2006predicting,
  title={Predicting the prognosis of breast cancer by integrating clinical and microarray data with {B}ayesian networks},
  author={Gevaert, Olivier and De Smet, Frank and Timmerman, Dirk and Moreau, Yves and De Moor, Bart},
  journal={Bioinformatics},
  volume={22},
  number={14},
  pages={e184--e190},
  year={2006}
}

@inproceedings{tsamardinos2003markov,
  title={Algorithms for large scale {M}arkov blanket discovery},
  author={Tsamardinos, Ioannis and Aliferis, Constantin F and Statnikov, Alexander},
  booktitle={Proceedings of the Florida AI Research Society (FLAIRS)},
  year={2003}
}

@article{aliferis2010local,
  title={Local causal and {M}arkov blanket induction for causal discovery and feature selection for classification},
  author={Aliferis, Constantin F and Statnikov, Alexander and Tsamardinos, Ioannis and Mani, Subramani and Koutsoukos, Xenofon D},
  journal={Journal of Machine Learning Research},
  volume={11},
  pages={171--234},
  year={2010}
}

@article{scutari2019learns,
  title={Who learns better {B}ayesian network structures: Accuracy and speed of structure learning algorithms},
  author={Scutari, Marco and Graafland, Catharina Elisabeth and Guti{\'e}rrez, Jos{\'e} Manuel},
  journal={International Journal of Approximate Reasoning},
  volume={115},
  pages={235--253},
  year={2019}
}

@article{cox1972regression,
  title={Regression models and life-tables},
  author={Cox, David R},
  journal={Journal of the Royal Statistical Society: Series B},
  volume={34},
  number={2},
  pages={187--220},
  year={1972}
}

@article{ishwaran2008rsf,
  title={Random survival forests},
  author={Ishwaran, Hemant and Kogalur, Udaya B and Blackstone, Eugene H and Lauer, Michael S},
  journal={The Annals of Applied Statistics},
  volume={2},
  number={3},
  pages={841--860},
  year={2008}
}

@article{hothorn2006survival,
  title={Survival ensembles},
  author={Hothorn, Torsten and B{\"u}hlmann, Peter and Dudoit, Sandrine and Molinaro, Annette and Van Der Laan, Mark J},
  journal={Biostatistics},
  volume={7},
  number={3},
  pages={355--373},
  year={2006}
}

@article{harrell1996multivariable,
  title={Multivariable prognostic models: Issues in developing models, evaluating assumptions and adequacy, and measuring and reducing errors},
  author={Harrell, Frank E and Lee, Kerry L and Mark, Daniel B},
  journal={Statistics in Medicine},
  volume={15},
  number={4},
  pages={361--387},
  year={1996}
}

@article{ang2010human,
  title={Human papillomavirus and survival of patients with oropharyngeal cancer},
  author={Ang, K Kian and Harris, Jonathan and Wheeler, Richard and Weber, Randal and Rosenthal, David I and Nguyen-T{\^a}n, Phuc Felix and Westra, William H and Chung, Christine H and Jordan, Richard C and Lu, Charles and others},
  journal={New England Journal of Medicine},
  volume={363},
  number={1},
  pages={24--35},
  year={2010}
}

@book{hernan2020causal,
  title={Causal Inference: What If},
  author={Hern{\'a}n, Miguel A and Robins, James M},
  publisher={Chapman \& Hall/CRC},
  address={Boca Raton},
  year={2020}
}

@article{kotevski2023machine,
  title={Machine learning and nomogram prognostic modeling for 2-year head and neck cancer--specific survival using electronic health record data: a multisite study},
  author={Kotevski, Damian P and Smee, Robert I and Vajdic, Claire M and Field, Matthew},
  journal={JCO Clinical Cancer Informatics},
  volume={7},
  pages={e2200128},
  year={2023},
  publisher={Wolters Kluwer Health}
}

@article{diamant2019deep,
  title={Deep learning in head \& neck cancer outcome prediction},
  author={Diamant, Andr{\'e} and Chatterjee, Avishek and Valli{\`e}res, Martin and Shenouda, George and Seuntjens, Jan},
  journal={Scientific reports},
  volume={9},
  number={1},
  pages={2764},
  year={2019},
  publisher={Nature Publishing Group UK London}
}

@article{lucas2004bayesian,
  title={Bayesian networks in biomedicine and health-care},
  author={Lucas, Peter J.F. and van der Gaag, Linda C. and Abu-Hanna, Ameen},
  journal={Artificial Intelligence in Medicine},
  volume={30},
  number={3},
  pages={201--214},
  year={2004},
  publisher={Elsevier}
}

@article{kyrimi2021comprehensive,
  title={A comprehensive scoping review of {B}ayesian networks in healthcare: Past, present and future},
  author={Kyrimi, Evangelia and McLachlan, Scott and Dube, Kudakwashe and Neves, Mariana R. and Fahmi, Ali and Fenton, Norman},
  journal={Artificial Intelligence in Medicine},
  volume={117},
  pages={102108},
  year={2021},
  publisher={Elsevier}
}

@article{mclachlan2020bayesian,
  title={Bayesian Networks in Healthcare: Distribution by Medical Condition},
  author={McLachlan, Scott and Dube, Kudakwashe and Hitman, Graham A. and Fenton, Norman E. and Kyrimi, Evangelia},
  journal={Artificial Intelligence in Medicine},
  volume={107},
  pages={101912},
  year={2020},
  publisher={Elsevier}
}

@inproceedings{rabinowicz2017prognostic,
  title={A prognostic model of glioblastoma multiforme using survival {B}ayesian networks},
  author={Rabinowicz, Simon and Hommersom, Arjen and Butz, Raphaela and Williams, Matt},
  booktitle={Conference on Artificial Intelligence in Medicine in Europe},
  volume={10259},
  pages={81--85},
  year={2017},
  organization={Springer},
  doi={10.1007/978-3-319-59758-4_9}
}

@article{kazmierski2023multi,
  title={Multi-institutional prognostic modeling in head and neck cancer: evaluating impact and generalizability of deep learning and radiomics},
  author={Kazmierski, Michal and Welch, Mattea and Kim, Sejin and McIntosh, Chris and Rey-McIntyre, Katrina and Huang, Shao Hui and Patel, Tirth and Tadic, Tony and Milosevic, Michael and Liu, Fei-Fei and others},
  journal={Cancer Research Communications},
  volume={3},
  number={6},
  pages={1140--1151},
  year={2023},
  publisher={American Association for Cancer Research}
}

@article{dorrich2025multimodal,
  title={A multimodal dataset for precision oncology in head and neck cancer},
  author={D{\"o}rrich, Marion and Balk, Matthias and Heusinger, Tatjana and Beyer, Sandra and Mirbagheri, Hamed and Fischer, David J and Kanso, Hassan and Matek, Christian and Hartmann, Arndt and Iro, Heinrich and others},
  journal={Nature Communications},
  volume={16},
  number={1},
  pages={7163},
  year={2025},
  publisher={Nature Publishing Group UK London}
}

@article{robins1986new,
  title={A new approach to causal inference in mortality studies with a sustained exposure period: Application to control of the healthy worker survivor effect},
  author={Robins, James},
  journal={Mathematical Modelling},
  volume={7},
  number={9-12},
  pages={1393--1512},
  year={1986}
}

@article{welch2024radcure,
  title={RADCURE: An open-source head and neck cancer CT dataset for clinical radiation therapy insights},
  author={Welch, Mattea L and Kim, Sejin and Hope, Andrew J and Huang, Shao Hui and Lu, Zhibin and Marsilla, Joseph and Kazmierski, Michal and Rey-McIntyre, Katrina and Patel, Tirth and O'Sullivan, Brian and others},
  journal={Medical Physics},
  volume={51},
  number={4},
  pages={3101--3109},
  year={2024},
  publisher={Wiley Online Library}
}

@article{heckerman1995learning,
  title={Learning Bayesian networks: The combination of knowledge and statistical data},
  author={Heckerman, David and Geiger, Dan and Chickering, David M},
  journal={Machine learning},
  volume={20},
  number={3},
  pages={197--243},
  year={1995},
  publisher={Springer}
}

@article{cooper1992bayesian,
  title={A Bayesian method for the induction of probabilistic networks from data},
  author={Cooper, Gregory F and Herskovits, Edward},
  journal={Machine learning},
  volume={9},
  number={4},
  pages={309--347},
  year={1992},
  publisher={Springer}
}

@book{spirtes2012causation,
  title={Causation, prediction, and search},
  author={Spirtes, Peter and Glymour, Clark and Scheines, Richard},
  volume={81},
  year={2012},
  publisher={Springer Science \& Business Media}
}

@article{yeo2000new,
  title={A new family of power transformations to improve normality or symmetry},
  author={Yeo, In-Kwon and Johnson, Richard A},
  journal={Biometrika},
  volume={87},
  number={4},
  pages={954--959},
  year={2000},
  publisher={Oxford University Press}
}

@article{ma2022association,
  title={Association of pack-years of cigarette smoking with survival and tumor progression among patients treated with chemoradiation for head and neck cancer},
  author={Ma, Sung Jun and Yu, Han and Yu, Brian and Waldman, Olivia and Khan, Michael and Chatterjee, Udit and Santhosh, Sharon and Gill, Jasmin and Iovoli, Austin J and Farrugia, Mark and others},
  journal={JAMA network open},
  volume={5},
  number={12},
  pages={e2245818},
  year={2022}
}

@misc{david2012survival,
  title={Survival analysis: a Self-Learning text},
  author={Kleinbaum, David G. and Klein, Mitchel},
  year={2012},
  publisher={Spinger}
}

@article{keil2014parametric,
  title={The parametric g-formula for time-to-event data: intuition and a worked example},
  author={Keil, Alexander P and Edwards, Jessie K and Richardson, David B and Naimi, Ashley I and Cole, Stephen R},
  journal={Epidemiology},
  volume={25},
  number={6},
  pages={889--897},
  year={2014},
  publisher={LWW}
}

@article{starke2023longitudinal,
  title={Longitudinal and multimodal radiomics models for head and neck cancer outcome prediction},
  author={Starke, Sebastian and Zwanenburg, Alexander and Leger, Karoline and Z{\"o}phel, Klaus and Kotzerke, J{\"o}rg and Krause, Mechthild and Baumann, Michael and Troost, Esther GC and L{\"o}ck, Steffen},
  journal={Cancers},
  volume={15},
  number={3},
  pages={673},
  year={2023},
  publisher={MDPI}
}

@article{chalker2022performance,
  title={Performance status ({PS}) as a predictor of poor response to immune checkpoint inhibitors ({ICI}) in recurrent/metastatic head and neck cancer ({RMHNSCC}) patients},
  author={Chalker, Cameron and Voutsinas, Jenna M and Wu, Qian Vicky and others},
  journal={Cancer Medicine},
  volume={11},
  number={24},
  pages={4815--4825},
  year={2022}
}

@article{prosnitz2005pretreatment,
  title={Pretreatment anemia is correlated with the reduced effectiveness of radiation and concurrent chemotherapy in advanced head and neck cancer},
  author={Yao, MS Bin and Farrell, Ph D Catherine L and Clough, BA Robert and Brizel, MD David M},
  journal={International Journal of Radiation Oncology* Biology* Physics},
  volume={61},
  number={4},
  pages={1087--1095},
  year={2005},
  publisher={Elsevier}
}

@article{ma2022defining,
  title={Defining the optimal threshold and prognostic utility of pre-treatment hemoglobin level as a biomarker for survival outcomes in head and neck cancer patients receiving chemoradiation},
  author={Ma, Sung Jun and Yu, Han and Khan, Michael and others},
  journal={Oral Oncology},
  volume={133},
  pages={106054},
  year={2022}
}

@article{roberts2016number,
  title={Number of positive nodes is superior to the lymph node ratio and A merican J oint C ommittee on C ancer N staging for the prognosis of surgically treated head and neck squamous cell carcinomas},
  author={Roberts, Thomas J and Colevas, A Dimitrios and Hara, Wendy and Holsinger, F Christopher and Oakley-Girvan, Ingrid and Divi, Vasu},
  journal={Cancer},
  volume={122},
  number={9},
  pages={1388--1397},
  year={2016},
  publisher={Wiley Online Library}
}

@article{lee2019number,
  title={Number of positive lymph nodes better predicts survival for oral cavity cancer},
  author={Lee, Hojun and others},
  journal={Journal of Surgical Oncology},
  volume={119},
  number={6},
  pages={675--682},
  year={2019}
}

@article{yang2018prognostic,
  title={Pretreatment neutrophil to lymphocyte ratio in determining the prognosis of head and neck cancer: a meta-analysis},
  author={Yu, Yalian and Wang, Hongbo and Yan, Aihui and Wang, Hailong and Li, Xinyao and Liu, Jiangtao and Li, Wei},
  journal={BMC cancer},
  volume={18},
  number={1},
  pages={383},
  year={2018},
  publisher={Springer}
}

@article{takenaka2018prognostic,
  title={Prognostic role of neutrophil-to-lymphocyte ratio in head and neck cancer: A meta-analysis},
  author={Takenaka, Yukinori and Oya, Ryohei and Kitamiura, Takahiro and others},
  journal={Head \& Neck},
  volume={40},
  number={3},
  pages={647--655},
  year={2018}
}

@article{mantel1966evaluation,
  title={Evaluation of survival data and two new rank order statistics arising in its consideration},
  author={Mantel, Nathan and others},
  journal={Cancer Chemother Rep},
  volume={50},
  number={3},
  pages={163--170},
  year={1966}
}

@article{dempster1977maximum,
  title={Maximum likelihood from incomplete data via the EM algorithm},
  author={Dempster, Arthur P and Laird, Nan M and Rubin, Donald B},
  journal={Journal of the royal statistical society: series B (methodological)},
  volume={39},
  number={1},
  pages={1--22},
  year={1977},
  publisher={Wiley Online Library}
}

@article{tibshirani1997lasso,
  title={The lasso method for variable selection in the Cox model},
  author={Tibshirani, Robert},
  journal={Statistics in medicine},
  volume={16},
  number={4},
  pages={385--395},
  year={1997},
  publisher={Wiley Online Library}
}

@article{benjamini1995controlling,
  title={Controlling the false discovery rate: a practical and powerful approach to multiple testing},
  author={Benjamini, Yoav and Hochberg, Yosef},
  journal={Journal of the Royal Statistical Society: Series B (Methodological)},
  volume={57},
  number={1},
  pages={289--300},
  year={1995},
  publisher={Wiley Online Library}
}

@article{grambsch1994proportional,
  title={Proportional hazards tests and diagnostics based on weighted residuals},
  author={Grambsch, Patricia M and Therneau, Terry M},
  journal={Biometrika},
  volume={81},
  number={3},
  pages={515--526},
  year={1994},
  publisher={Oxford University Press}
}

@article{schoenfeld1982partial,
  title={Partial residuals for the proportional hazards regression model},
  author={Schoenfeld, David},
  journal={Biometrika},
  volume={69},
  number={1},
  pages={239--241},
  year={1982},
  publisher={Oxford University Press}
}

@article{hoff2012importance,
  title={Importance of hemoglobin concentration and its modification for the outcome of head and neck cancer patients treated with radiotherapy},
  author={Hoff, Camilla Molich},
  journal={Acta Oncologica},
  volume={51},
  number={4},
  pages={419--432},
  year={2012},
  publisher={Taylor \& Francis}
}

\newpage
\appendix

\section{SA-BN Algorithm Details}
\label{app:algo_detail}

\paragraph{Survival scoring for target--parent edges.}
Let $(t_i, \delta_i)$ denote the observed time and event indicator for patient $i$, and let $\mathbf{x}_i$ be the feature vector. For a candidate parent set $\mathrm{Pa}$ of the survival node, the Cox partial log-likelihood is:
\begin{equation}
  \ell(\boldsymbol{\beta}; \mathrm{Pa}) \;=\; \sum_{i:\,\delta_i=1} \!\Bigl[\, \boldsymbol{\beta}^\top \mathbf{x}_{i,\mathrm{Pa}} \;-\; \log\!\!\sum_{j \in \mathcal{R}(t_i)} \!\exp\bigl(\boldsymbol{\beta}^\top \mathbf{x}_{j,\mathrm{Pa}}\bigr)\,\Bigr]
  \label{eq:cox_pll}
\end{equation}
where $\mathcal{R}(t_i)$ is the risk set and $\mathbf{x}_{i,\mathrm{Pa}}$ is the subvector of features in $\mathrm{Pa}$. Starting from an empty parent set with the null-model log-likelihood $\ell_0 = -\sum_{i:\delta_i=1} \log |\mathcal{R}(t_i)|$, at each iteration we add the candidate $X_k$ whose inclusion yields the smallest likelihood-ratio $p$-value, provided the test $\Lambda = 2[\ell(\hat{\boldsymbol{\beta}}; \mathrm{Pa} \cup \{X_k\}) - \ell(\hat{\boldsymbol{\beta}}; \mathrm{Pa})] \sim \chi^2_{\text{df}}$ satisfies $p < 0.05$; selection stops when no remaining feature meets this threshold or when the maximum in-degree (5) is reached. Cox models are fit using the Breslow approximation with a small $L_2$ penalty ($\lambda = 0.01$) for numerical stability~\citep{cox1972regression}. All LRT comparisons within a single selection run use the same complete-case subset so that nested models are validly comparable.

We instantiate this survival-aware scoring within three structurally different BN algorithms and take a consensus ($\geq$2/3) of their outputs, reducing sensitivity to any single search strategy. \\
i) SA-HC/Cox: Hill-Climbing with BDeu scoring ~\citep{heckerman1995learning} (equivalent sample size 10, max in-degree 5) for feature--feature edges; target--parent edges selected by forward Cox LRT as described above ($p < 0.05$). \\
ii) SA-HC/LR: Hill-Climbing with K2 scoring \citep{cooper1992bayesian} for feature--feature edges; target--parent edges selected by forward conditional multi-group log-rank test~\citep{mantel1966evaluation} ($p < 0.01$), where at each step the candidate $X_k$ is tested for conditional independence from survival given the previously selected parents. Continuous candidates are stratified using the same GMM discretization as feature--feature scoring. \\
iii) SA-PC/LR: The PC (Peter--Clark) constraint-based algorithm~\citep{spirtes2012causation} with conditioning-set size up to 3 and significance level $\alpha = 0.05$. Feature--feature edges use $\chi^2$ conditional independence tests on discretized strata; feature--target edges use a stratified log-rank test where the conditioning set defines the strata, and strata with fewer than 10 observations are skipped. The stratified $p$-value combines per-stratum log-rank statistics by summing $\chi^2$ values and degrees of freedom. \\[\baselineskip]
Continuous features are discretized via Yeo--Johnson normalization~\citep{yeo2000new} followed by Gaussian mixture clustering, with the number of components selected by AIC from $\{2, \ldots, \min(6,\, \lfloor N/15\rfloor + 1)\}$ per feature~\citep{dempster1977maximum}. All transformations are fit only to the training split; test-set values are transformed using the fitted parameters. Discretization is applied only to BDeu/K2 feature--feature scoring and to the log-rank tests; the Cox partial log-likelihood for feature--target edges operates on the original continuous values. The survival Markov blanket is identified by three methods applied to the training split and combined by consensus ($\geq$2/3): i) DAG-derived: parents, children, and co-parents of the survival node in the SA-BN consensus graph. ii) Cox stepwise: forward selection with LRT $p<0.05$, followed by backward pruning at $p>0.10$, both using the Cox partial log-likelihood on the complete-case training subset with $L_2$ penalizer $0.01$. iii) Univariate C-index top-$k$: features ranked by Harrell's C-index~\citep{harrell1996multivariable} on the training split; the top $k$ are selected, where $k = \max\bigl(|\mathrm{MB}_{\text{DAG}}|,\, |\mathrm{MB}_{\text{Cox}}|,\, 5\bigr)$. A feature enters the consensus Markov blanket if selected by at least 2 of these 3 methods. The resulting blanket is stable across LRT thresholds $p \in [0.01, 0.10]$ and BDeu equivalent sample sizes $\in [1, 50]$. Markov blanket stability (Figure~\ref{fig:mb_stability})  is assessed by 50 bootstrap resamples of the training split, drawn with replacement at the original sample size. Each resample runs through the full structure-learning, consensus, and MB-discovery pipeline; a feature's stability is the fraction of resamples in which it enters the consensus Markov blanket. The 70\% threshold in Figure~\ref{fig:mb_stability} corresponds to selection in at least 35 of 50 resamples. All structure learning, MB discovery, and discretization fitting use only the training split; the temporally held-out test set is reserved for final evaluation of the survival model.

\section{BN Structure, Bootstrap Stability and Survival Modeling}
\label{app:bn_figs}
Figures~\ref{fig:bn_rad} and~\ref{fig:bn_han} show the SA-BN consensus structures for the radiation and surgical cohorts, respectively. Nodes are colored by their role relative to the survival target, and edges reflect consensus structure across bootstrap resamples. Table~\ref{tab:survival_models}
compares Cox PH, RSF, and GBS across feature sets.
The SA-BN MB retains $>$95\% of all-feature performance in both cohorts (radiation: RSF 0.801 vs.\ 0.803; surgical: RSF 0.726 vs.\ 0.758), while non-MB features alone drop to C\,=\,0.685 in the radiation cohort. The small gap between MB and all-feature models is expected; the MB is designed to be the minimal \emph{sufficient} set that captures the outcome's conditional independence structure, not the set that maximizes predictive performance.     Its value is interpretive as it answers ``which features matter and why'' rather than ``what is the highest achievable C-index''.

\begin{table}[ht]
\centering
\caption{Survival model C-index [95\% CI]. Bold: best per cohort per feature set.}
\label{tab:survival_models}
\begin{tabular}{@{}llll@{}}
\toprule
\textbf{Features} & \textbf{Model} & \textbf{Radiation} & \textbf{Surgical} \\
\midrule
\multirow{3}{*}{All (13/15)}
  & Cox PH & 0.792 [.75-.83] & 0.686 [.57-.80] \\
  & RSF    & 0.803 [.77-.84] & \textbf{0.758} [.67-.83] \\
  & GBS    & \textbf{0.806} [.77-.84] & 0.681 [.59-.78] \\
\midrule
\multirow{3}{*}{SA-MB (9/5)}
  & Cox PH & 0.791 [.75-.83] & 0.625 [.51-.74] \\
  & RSF    & \textbf{0.801} [.76-.84] & \textbf{0.726} [.64-.81] \\
  & GBS    & 0.798 [.76-.83] & 0.672 [.57-.77] \\
\midrule
\multirow{2}{*}{Non-MB}
  & Cox PH & 0.681 [.63-.73] & 0.681 [.59-.77] \\
  & RSF    & 0.685 [.64-.73] & 0.736 [.65-.81] \\
\bottomrule
\end{tabular}
\end{table}

\section{Risk Stratification}
\label{app:km}

\begin{figure}[ht]
\centering
\subfigure[Radiation Cohort.]{\includegraphics[width=0.48\linewidth]{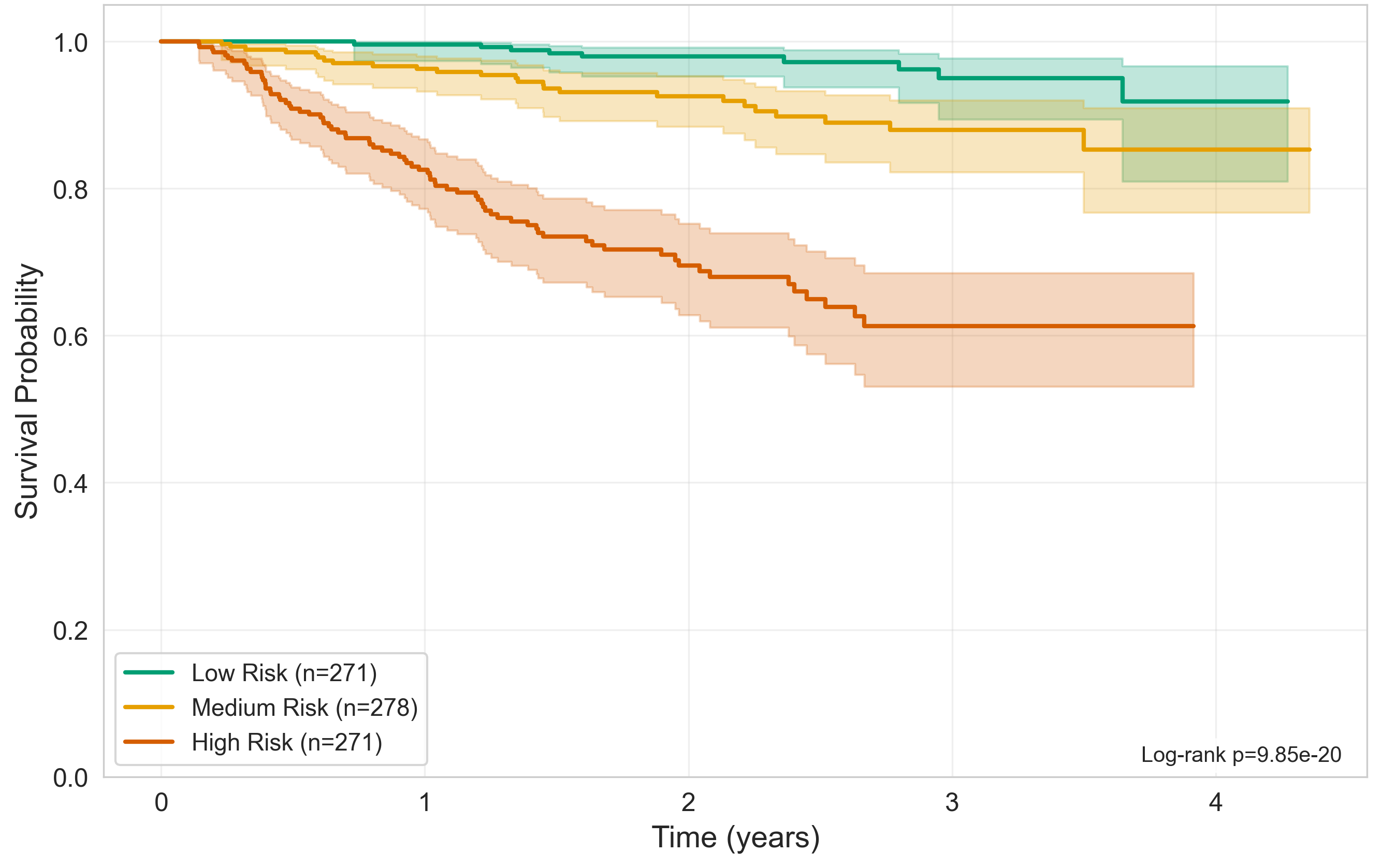}}
\subfigure[Surgical Cohort.]{\includegraphics[width=0.48\linewidth]{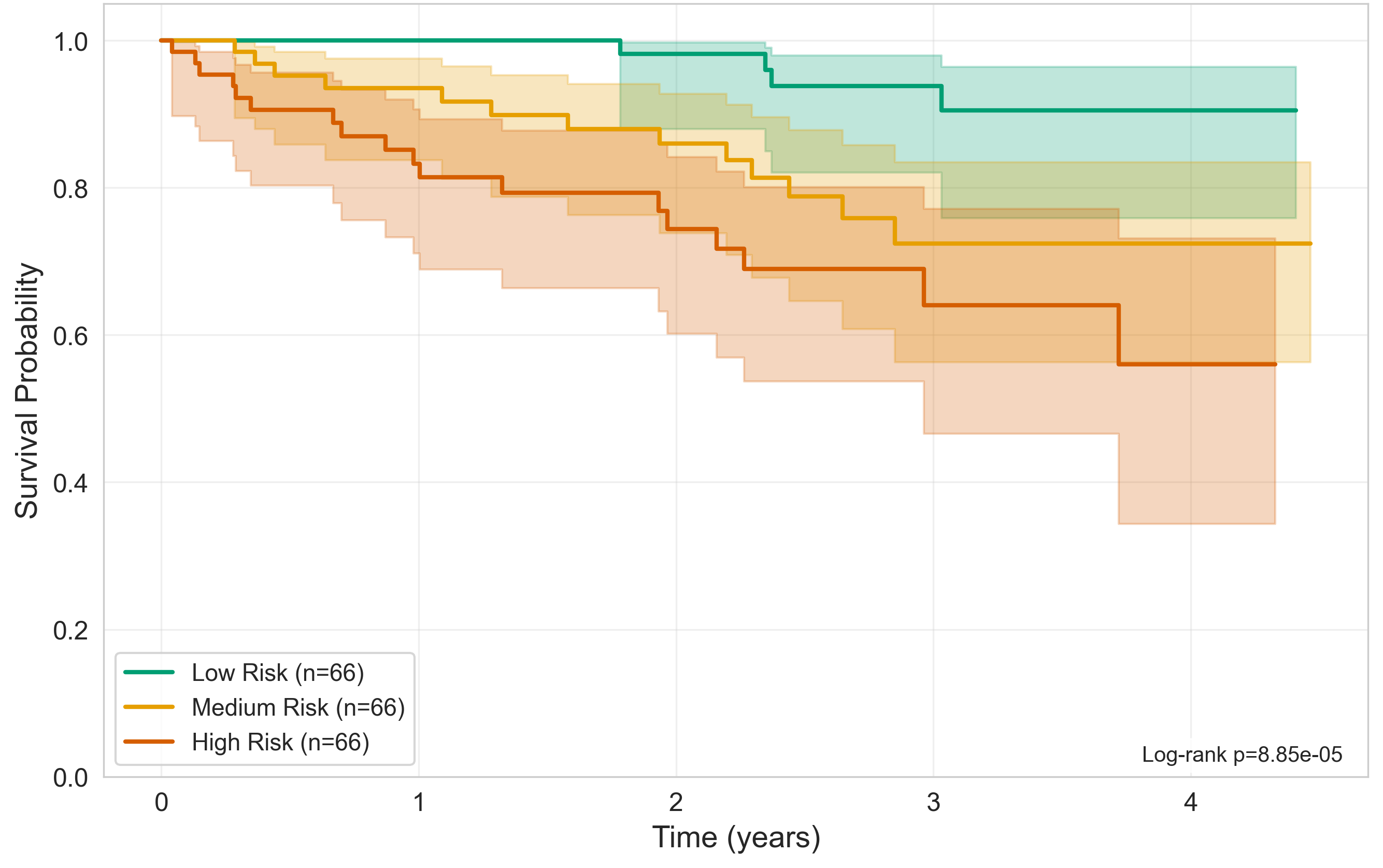}}
\caption{KM survival curves stratified by RSF risk score tertiles using SA-BN MB features.}
\label{fig:km_strata}
\end{figure}

We stratify patients into tertiles based on RSF-predicted risk scores using SA-MB features.
In both cohorts, the resulting Kaplan--Meier curves are well separated (Figure~\ref{fig:km_strata}).
In the radiation cohort, the high-risk tertile has approximately 40\% survival at 4~years, compared with over 90\% in the low-risk tertile.
In the surgical cohort, separation is smaller but still significant ($p < 0.001$), reflecting the wider confidence intervals from the smaller test set ($n{=}198$, 35 events).

\section{Ablation: Feature Selection by Method}
\label{app:ablation_detail}

Table~\ref{tab:ablation} compares feature selection across three conditions: binary-proxy BN (evaluable patients only), SA-BN on evaluable patients only (same patients as binary, but with Cox scoring), and SA-BN on all patients (Cox scoring + censored patients included).
In the radiation cohort, SA-BN on evaluable patients recovers 8 of 9 features, confirming that the Cox scoring formulation is the primary driver; the extra 359 censored patients add only overall stage.
In the surgical cohort, SA-BN on evaluable patients recovers all 5 features (plus T$\times$N).
Features not selected by any method (sex, N~stage, Histological Subtype on radiation; site, pT, grade, HPV, sex, smoking on surgical) are consistently excluded, confirming that the MB differences are not random.

\section{G-Computation Details}
\label{app:gcomp}

For a feature $X_k$ at level $x$, the interventional RMST is:
\begin{equation}
  \mathrm{RMST}(\mathrm{do}(X_k{=}x)) \;=\; \int_0^{\tau} \frac{1}{n}\sum_{i=1}^{n} \hat{S}(t \mid X_k{=}x,\, \mathbf{C}_i)\,dt
\end{equation}
where $\tau{=}5$~years, $\hat{S}$ is the Cox survival function, and $\mathbf{C}_i$ are the other Markov blanket parents at patient $i$'s actual values.
Setting $X_k{=}x$ for all patients while keeping $\mathbf{C}_i$ fixed implements the do-operation; averaging marginalizes out confounders~\citep{robins1986new,hernan2020causal}. \citet{shah2026bayesian} use do-calculus on the BN's conditional probability tables for the same purpose, which requires manually reversing edges from the outcome to its predictors before the intervention can be computed.
G-computation avoids this by using the BN to identify the adjustment set and delegating causal estimation to a Cox model that does not depend on edge directions. The key assumption is that the MB parents constitute a sufficient adjustment set for the backdoor criterion, which is plausible but not guaranteed from observational data~\citep{hernan2020causal}.
Bootstrap 95\% CIs from 50 resamples are reported.

\section{Causal Analysis: Hazard Ratios and Interventional RMST}
\label{app:causal_figs}

\begin{figure}[ht]
\centering
\subfigure[Radiation Cohort]{\includegraphics[height=4.2cm]{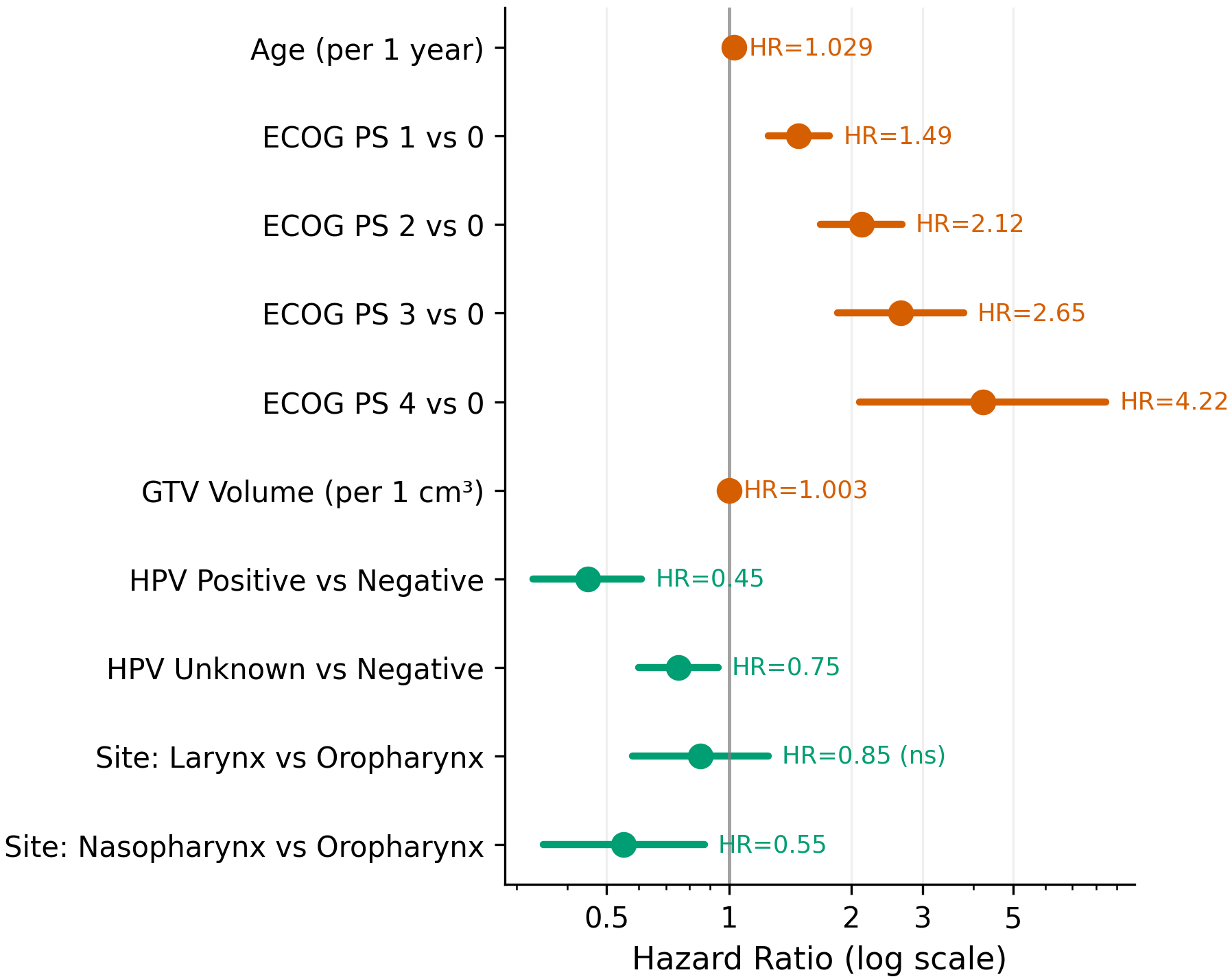}}
\subfigure[Surgical Cohort]{\includegraphics[height=4.2cm]{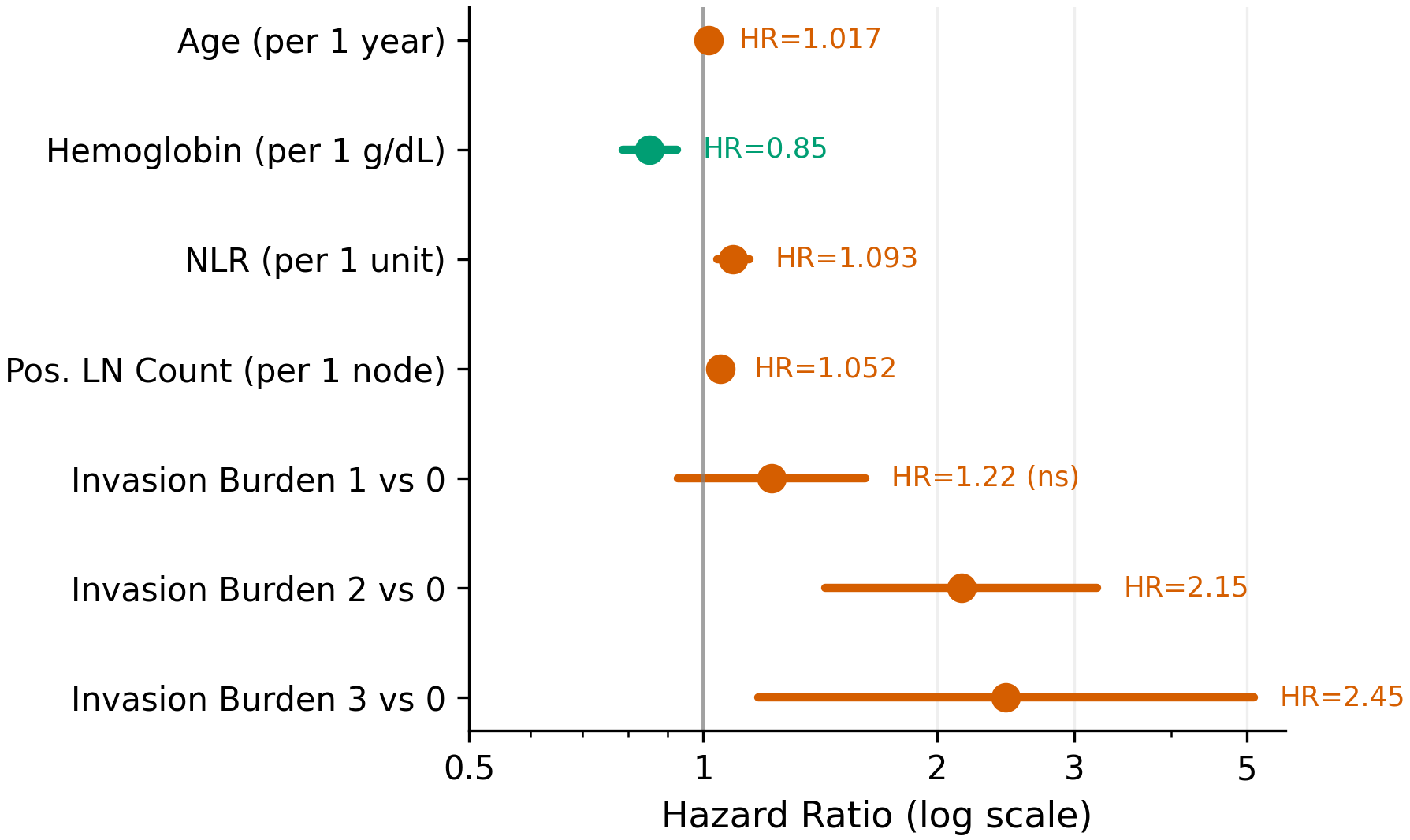}}
\caption{Causal hazard ratio forest plots with 95\% CIs on log scale. DAG-informed covariate adjustment.}
\label{fig:hazard_ratios}
\end{figure}

\begin{figure}[ht]
\centering
\includegraphics[width=0.95\linewidth]{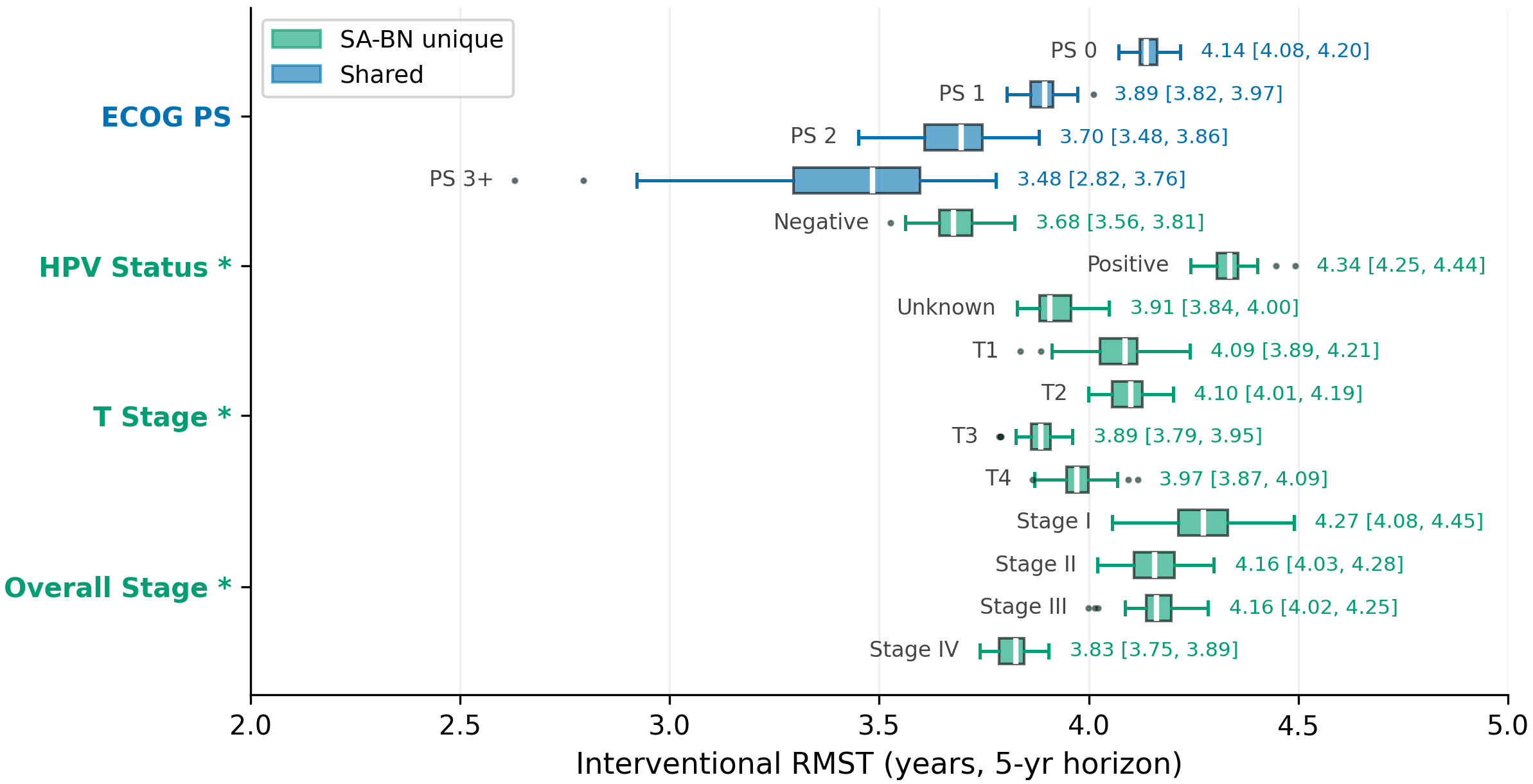}
\caption{Interventional RMST (5-year horizon) on the radiation cohort by feature level estimated via G-computation. Green = SA-BN unique; blue = shared with binary Markov blanket.}
\label{fig:rmst_levels_rad}
\end{figure}

\begin{figure}[ht]
\centering
\includegraphics[width=0.8\linewidth]{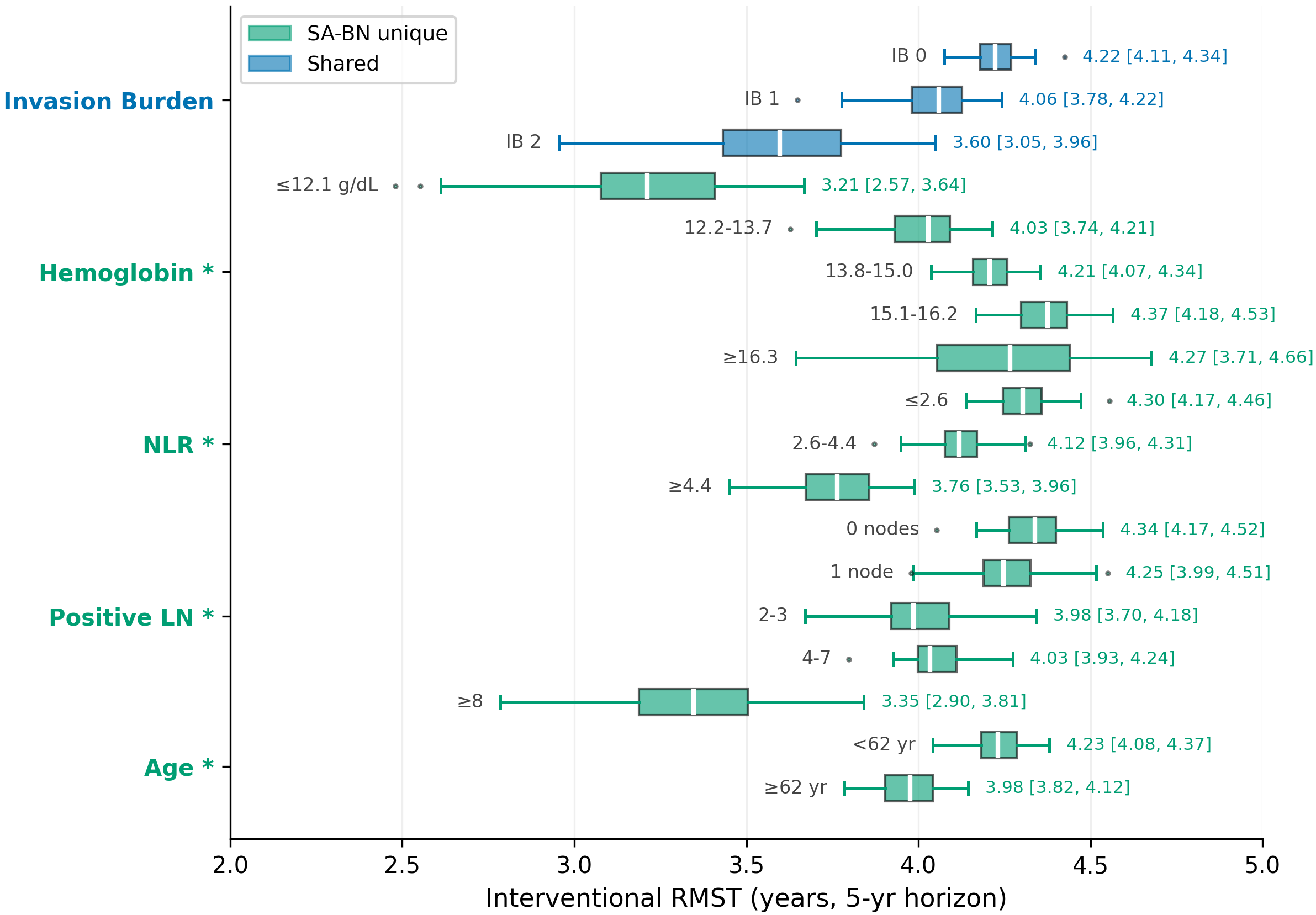}
\caption{\normalfont Interventional RMST (5-year horizon) on the surgical cohort by feature level estimated via G-computation.}
\label{fig:rmst_levels_han}
\end{figure}

The hazard ratios for the primary SA-BN exposures are shown in Table~\ref{tab:hazard_ratios} and Figure~\ref{fig:hazard_ratios}, complementing the interventional RMST figures (Figures~\ref{fig:rmst_levels_rad} and \ref{fig:rmst_levels_han}) with per-unit effect sizes. 
In the radiation cohort, ECOG performance status exhibits a monotonic hazard gradient from PS~1 (HR\,=\,1.49) to PS~4 (HR\,=\,4.22), supporting its selection as a direct survival parent.
In the surgical cohort, invasion burden shows a threshold effect at IB$\geq$2, with hemoglobin as a continuous risk factor.

\begin{table}[ht]
\centering
\caption{Causal hazard ratios from DAG-informed Cox PH. $^*p < 0.05$.}
\label{tab:hazard_ratios}
\begin{tabular}{@{}llrl@{}}
\toprule
\textbf{Cohort} & \textbf{Variable} & \textbf{HR} & \textbf{$p$} \\
\midrule
Radiation & ECOG PS 1 vs.\ 0 & 1.485 & $< 0.001^*$ \\
Radiation & ECOG PS 2 vs.\ 0 & 2.120 & $< 0.001^*$ \\
Radiation & ECOG PS 3 vs.\ 0 & 2.646 & $< 0.001^*$ \\
Radiation & ECOG PS 4 vs.\ 0 & 4.216 & $< 0.001^*$ \\
\midrule
Surgical & Age (per year) & 1.017 & $0.005^*$ \\
Surgical & Hemoglobin (per g/dL) & 0.853 & $< 0.001^*$ \\
Surgical & NLR (per unit) & 1.093 & $< 0.001^*$ \\
Surgical & Positive \ LN (per node) & 1.052 & $< 0.001^*$ \\
Surgical & IB 1 vs.\ 0 & 1.224 & $0.153$ \\
Surgical & IB 2 vs.\ 0 & 2.148 & $< 0.001^*$ \\
Surgical & IB 3 vs.\ 0 & 2.452 & $0.017^*$ \\
\bottomrule
\end{tabular}
\end{table}

\section{Proportional-Hazards Testing}
\label{app:ph}

We assess the proportional hazards (PH) assumption of the DAG-informed Cox models using Schoenfeld residual tests, applied per feature within each cohort's survival Markov blanket, and corrected for multiple comparisons with the Benjamini--Hochberg (BH) FDR procedure (Table~\ref{tab:ph}). After correction, the only violation is overall stage in the radiation cohort ($q = 7.2\times10^{-5}$); no surgical Markov-blanket feature violates PH ($q \ge 0.17$). As a sensitivity analysis, we refit the radiation Cox model stratified on overall stage: all hazard ratios retain direction and magnitude (ECOG PS $1.95\to1.82$, HPV $0.47\to0.43$, site $0.34\to0.42$, T~stage $0.69\to0.77$, treatment modality $1.61\to1.76$), so the hazard-ratio and interventional-RMST interpretations are unaffected.

\begin{table}[ht]
\centering
\footnotesize
\caption{Schoenfeld PH residual tests with BH-FDR within each cohort's survival Markov blanket. Bold $q<0.05$ indicates a PH violation.}
\label{tab:ph}
\begin{tabular}{@{}llrrr@{}}
\toprule
\textbf{Cohort} & \textbf{Feature} & \textbf{stat} & \textbf{$p_{\text{raw}}$} & \textbf{$q_{\text{BH}}$} \\
\midrule
Radiation & overall stage        & 19.9  & $8\times10^{-6}$ & $7.2\times10^{-5}$ \\
Radiation & smoking (pack-yr)     & 5.64  & 0.018 & 0.079 \\
Radiation & age                  & 3.67  & 0.055 & 0.107 \\
Radiation & site                 & 3.57  & 0.059 & 0.107 \\
Radiation & GTV                  & 3.37  & 0.066 & 0.107 \\
Radiation & T stage              & 3.26  & 0.071 & 0.107 \\
Radiation & HPV status           & 1.61  & 0.205 & 0.264 \\
Radiation & treatment modality   & 1.38  & 0.241 & 0.271 \\
Radiation & ECOG PS              & 1.09  & 0.296 & 0.296 \\
\midrule
Surgical  & blood hemoglobin     & 4.50  & 0.034 & 0.169 \\
Surgical  & invasion burden      & 3.19  & 0.074 & 0.185 \\
Surgical  & positive LN count    & 2.35  & 0.126 & 0.209 \\
Surgical  & NLR                  & 0.54  & 0.464 & 0.580 \\
Surgical  & age                  & 0.25  & 0.616 & 0.616 \\
\bottomrule
\end{tabular}
\end{table}

\section{Quantitative Markov-Blanket Membership}
\label{app:quant_mb}

For each survival Markov-blanket feature, its 50-bootstrap selection frequency, DAG-informed hazard ratio, BH-FDR-corrected Cox $p$-value ($q$), and the bootstrap stability of its directed edge into the survival node are reported in Table~\ref{tab:quant_mb}. Every feature survives multiple-comparison correction at $q<0.05$ except radiation T~stage ($q=0.42$); in the surgical cohort, all five features survive. Directed-edge-into-survival stabilities from the same 50 bootstraps confirm that the highest-frequency features also anchor the graph: in the radiation cohort, ECOG PS enters survival in 100\% of resamples, age in 98\%, and site in 96\%; in the surgical cohort, hemoglobin in 82\% and invasion burden in 74\%.

\begin{table}[ht]
\centering
\footnotesize
\caption{Survival Markov-blanket membership: 50-bootstrap selection frequency, hazard ratio (DAG-informed Cox) and BH-FDR $q$-value.}
\label{tab:quant_mb}
\begin{tabular}{@{}llrrr@{}}
\toprule
\textbf{Cohort} & \textbf{Feature} & \textbf{Sel.\ freq.} & \textbf{HR} & \textbf{$q_{\text{BH}}$} \\
\midrule
\multirow{9}{*}{Radiation}
 & HPV status         & 100\% & 0.43 & $2.5\times10^{-13}$ \\
 & age                & 100\% & 1.03 & $2.0\times10^{-12}$ \\
 & treatment modality & 100\% & 1.84 & $6.0\times10^{-10}$ \\
 & GTV                & 100\% & 1.01 & $4.3\times10^{-8}$ \\
 & ECOG PS            & 100\% & 1.83 & $1.3\times10^{-7}$ \\
 & smoking (pack-yr)  & 100\% & 1.01 & $2.3\times10^{-7}$ \\
 & site               & 100\% & 0.41 & $6.9\times10^{-6}$ \\
 & overall stage      & 68\%  & 2.09 & $3.3\times10^{-4}$ \\
 & T stage            & 66\%  & 0.81 & 0.42 \\
\midrule
\multirow{5}{*}{Surgical}
 & blood hemoglobin   & 94\%  & 0.82 & $4.8\times10^{-4}$ \\
 & positive LN count  & 88\%  & 1.06 & $7.5\times10^{-5}$ \\
 & age                & 88\%  & 1.03 & $9.0\times10^{-5}$ \\
 & invasion burden    & 86\%  & 2.87 & $4.2\times10^{-5}$ \\
 & NLR                & 78\%  & 1.08 & $6.9\times10^{-3}$ \\
\bottomrule
\end{tabular}
\end{table}

\section{Extended Downstream-Model Comparison}
\label{app:model_perf}

We compare four survival models: Cox PH, RSF, gradient-boosted survival (GBS), and LASSO-penalized Cox, holding the model fixed and varying only the feature set (survival-aware vs.\ binary Markov blanket), across three metrics (Table~\ref{tab:model_r1}). In the radiation cohort, the survival-aware set wins every model and metric. In the surgical cohort, it wins for the non-linear RSF and on the integrated Brier score across all models, while a linear Cox or LASSO-Cox narrowly favors the binary set on discrimination, an observation consistent with non-linear models transferring better on the smaller surgical cohort.

\begin{table}[ht]
\centering
\footnotesize
\setlength{\tabcolsep}{4pt}
\caption{Downstream survival models on the survival-aware vs.\ binary Markov blanket. Each cell is \textbf{SA-MB} / binary-MB; Harrell C and tAUC higher is better, IBS lower is better; bold marks the winning feature set. Values are from the extended multi-model benchmark, whose feature encoding/preprocessing differs slightly from the primary run behind the Table~\ref{tab:sabn_comparison} headline. Radiation values and the survival-aware surgical values agree with the primary run to within bootstrap noise; the surgical binary-MB C-index is higher here (RSF 0.678 vs.\ the primary-run 0.648), so the survival-aware surgical gain under this benchmark ($+0.043$ for RSF) is smaller than the headline $+0.078$ while remaining in the same direction.}
\label{tab:model_r1}
\begin{tabular}{@{}llccc@{}}
\toprule
\textbf{Cohort} & \textbf{Model} & \textbf{Harrell C} & \textbf{tAUC} & \textbf{IBS$\downarrow$} \\
\midrule
\multirow{4}{*}{Radiation}
 & Cox PH & \textbf{.792}/.766 & \textbf{.840}/.812 & \textbf{.067}/.070 \\
 & RSF    & \textbf{.804}/.770 & \textbf{.848}/.815 & \textbf{.069}/.072 \\
 & GBS    & \textbf{.799}/.766 & \textbf{.844}/.811 & \textbf{.067}/.071 \\
 & LASSO-Cox & \textbf{.790}/.757 & \textbf{.837}/.802 & \textbf{.069}/.073 \\
\midrule
\multirow{4}{*}{Surgical}
 & Cox PH & .625/\textbf{.639} & .671/\textbf{.695} & \textbf{.093}/.096 \\
 & RSF    & \textbf{.721}/.678 & \textbf{.767}/.722 & \textbf{.089}/.096 \\
 & GBS    & \textbf{.672}/.663 & .700/\textbf{.714} & \textbf{.094}/.099 \\
 & LASSO-Cox & .636/\textbf{.686} & .680/\textbf{.730} & \textbf{.093}/.095 \\
\bottomrule
\end{tabular}
\end{table}

\section{Parity with Survival-Native Selectors}
\label{app:parity}

We compare the survival Markov blanket against size-matched LASSO-Cox and RSF variable-importance feature sets, with a downstream Cox PH model fit on each selected set for a like-for-like comparison (Table~\ref{tab:parity}). A paired-bootstrap C-index difference test ($B=1000$) with BH-FDR does not distinguish SA-BN from a size-matched LASSO-Cox or RSF-VI set on either cohort (all $q>0.5$). The only significant difference is a larger cross-validated 12-feature LASSO-Cox model, which exceeds the 6-feature radiation blanket ($q=0.04$). The advantage is attributable to the larger feature set at a fixed estimator.

\begin{table}[ht]
\centering
\footnotesize
\caption{Feature-selection parity. Test-set C-index (Cox PH on each selected set) and paired-bootstrap BH-FDR $q$-value versus the SA-BN blanket. ``--'' = not applicable.}
\label{tab:parity}
\begin{tabular}{@{}lccrr@{}}
\toprule
\textbf{Selection method} & \textbf{$k$ (rad/surg)} & \textbf{Rad.\ C} & \textbf{Surg.\ C} & \textbf{$q$ vs.\ SA-BN (rad/surg)} \\
\midrule
SA-BN blanket (Cox)    & 9 / 5 & 0.791 & 0.725 & -- \\
LASSO-Cox (size-matched) & 6 / 4 & 0.769 & 0.699 & 0.64 / 0.57 \\
LASSO-Cox (CV-best)    & 12 / -- & 0.795 & -- & \textbf{0.04} / -- \\
RSF-VI (top-$k$)       & 6 / 5 & 0.769 & 0.674 & 0.64 / 0.57 \\
\bottomrule
\end{tabular}
\end{table}

\section{Computational Cost}
\label{app:runtime}

 We report wall-clock times on a single CPU thread. A single ensemble pass is dominated by the PC constraint-based step at large $n$ (Table~\ref{tab:runtime}). The 50-bootstrap stability analysis is a one-time offline step and is not required at deployment. A single pass at $n\approx100$ ($\sim$2.5~s) is a fast route for privacy- or size-constrained cohorts, and a single SA-BN pass remains cheaper than RSF permutation importance.

\begin{table}[ht]
\centering
\footnotesize
\caption{Runtime, single CPU thread. Reference selectors (RSF variable importance, LASSO-Cox) shown for comparison.}
\label{tab:runtime}
\begin{tabular}{@{}lrr@{}}
\toprule
\textbf{Operation} & \textbf{Radiation} & \textbf{Surgical} \\
\midrule
1 ensemble pass                & $\sim$137 s & $\sim$13 s \\
single pass, $n\approx100$     & $\sim$2.5 s & $\sim$2.5 s \\
50-bootstrap stability (offline) & $\sim$114 min & $\sim$11 min \\
RSF-VI (reference)             & $\sim$16.6 min & -- \\
LASSO-Cox (reference)          & $<$1 s & $<$1 s \\
\bottomrule
\end{tabular}
\end{table}

\end{document}